\documentstyle[psfig,epsfig,graphicx,amsfonts,url]{mn}

\newcommand{\be}{\begin{equation}}
\newcommand{\ee}{\end{equation}}
\newcommand{\ba}{\begin{eqnarray}}
\newcommand{\ea}{\end{eqnarray}}
\newcommand{\nn}{\nonumber \\}

\newcommand{\idot}{\mbox{${\rm i}$}}

\newcommand{\bell}{{\mbox{\boldmath{$\ell$}}}}

\newcommand{\mnras}{MNRAS}
\newcommand{\prd}{Phys. Rev. D}
\newcommand{\apjl}{Astro. Phys. Journal Letters}
\newcommand{\apj}{Astro. Phys. Journal}
\newcommand{\aap}{AAP}
\newcommand{\apjs}{Astro. Phys. Journal Supplements}
\newcommand{\jcap}{Journal of Cosmology and Astroparticle Physics}

\def\gs{\mathrel{\raise1.16pt\hbox{$>$}\kern-7.0pt %
\lower3.06pt\hbox{{$\scriptstyle \sim$}}}}         %
\def\ls{\mathrel{\raise1.16pt\hbox{$<$}\kern-7.0pt %
\lower3.06pt\hbox{{$\scriptstyle \sim$}}}}         %

\title[3D Cosmic Shear]{3D Cosmic Shear: Cosmology from CFHTLenS}
\author[Kitching et al.] 
{T. D. Kitching\thanks{t.kitching@ucl.ac.uk}$^{1}$,  
A. F. Heavens$^{2}$, J. Alsing$^{2}$, T. Erben$^{3}$, C. Heymans$^{4}$, 
\newauthor
H. Hildebrandt$^{3,5}$, H. Hoekstra$^{6}$, A. Jaffe$^{2}$, A. Kiessling$^{7}$, Y. Mellier$^{8,9}$, L. Miller$^{10}$, 
\newauthor
L. van Waerbeke$^{5}$, J. Benjamin$^{5}$, J. Coupon$^{11}$, L. Fu$^{12}$, M. J. Hudson$^{13,14}$, 
\newauthor
M. Kilbinger$^{9}$, K. Kuijken$^{6}$, B. T. P. Rowe$^{15}$, T. Schrabback$^{3,6,16}$, 
\newauthor 
E. Semboloni$^{6}$, M. Velander$^{6,10}$
\\
$^{1}$Mullard Space Science Laboratory, University College London, Holmbury St Mary, Dorking, Surrey RH5 6NT, UK\\
$^{2}$Imperial Centre for Inference and Cosmology, Imperial College London, Prince Consort Road, London SW7 2AZ, U.K.\\
$^{3}$Argelander Institute for Astronomy, University of Bonn, Auf dem H{\"u}gel 71, 53121 Bonn, Germany\\
$^{4}$SUPA, Institute for Astronomy, University of Edinburgh, Royal Observatory, Blackford Hill, Edinburgh, EH9 3HJ, UK.\\
$^{5}$University of British Columbia, Department of Physics and Astronomy, 6224 Agricultural Road, Vancouver, B.C. V6T 1Z1, Canada\\
$^{6}$Leiden Observatory, Leiden University, Niels Bohrweg 2, 2333 CA Leiden, The Netherlands\\
$^{7}$Jet Propulsion Laboratory, California Institute of Technology, 4800 Oak Grove Drive, Pasadena, CA 91109, USA\\
$^{8}$Institut d'Astrophysique de Paris, CNRS, UMR 7095, 98 bis Boulevard Arago, F-75014 Paris, France\\
$^{9}$CEA/Irfu/SAp Saclay, Laboratoire AIM, 91191 Gif-sur-Yvette, France\\
$^{10}$University of Oxford, Denys Wilkinson Building, Department of Physics, Wilkinson Building, Keble Road, Oxford OX1 3RH, U.K.\\
$^{11}$Astronomical Observatory of the University of Geneva, ch. d'Ecogia 16, 1290 Versoix, Switzerland\\
$^{12}$Shanghai Key Lab for Astrophysics, Shanghai Normal University, 100 Guilin Road, 200234, Shanghai, China\\  
$^{13}$Dept. of Physics and Astronomy, University of Waterloo, Waterloo, ON, N2L 3G1, Canada\\
$^{14}$Perimeter Institute for Theoretical Physics, 31 Caroline Street N, Waterloo, ON, N2L 1Y5, Canada\\
$^{15}$Department of Physics and Astronomy, University College London,Gower Street, London WC1E 6BT, U.K.\\
$^{16}$Kavli Institute for Particle Astrophysics and Cosmology, Stanford University, 382 Via Pueblo Mall, Stanford, CA 94305-4060, USA}

\date{}

\pagerange{\pageref{firstpage}--\pageref{lastpage}}

\pubyear{2013}

\begin{document}

\maketitle

\label{firstpage}


\begin{abstract}
\noindent 
This paper presents the first application of 3D cosmic shear to a wide-field weak lensing survey. 3D cosmic shear is 
a technique that analyses weak lensing in three dimensions using a spherical harmonic approach, and does not bin data 
in the redshift direction. This is applied to CFHTLenS, a $154$ square degree imaging survey with a median redshift of $0.7$ 
and an effective number density of $11$ galaxies per square arcminute usable for weak lensing. To account for survey masks 
we apply a 3D pseudo-$C_{\ell}$ approach on weak lensing data, and to avoid uncertainties in the highly non-linear regime, 
we separately analyse radial wavenumbers $k\leq 1.5h$ Mpc$^{-1}$ and $k\leq 5.0h$ Mpc$^{-1}$, and angular wavenumbers 
$\ell\approx 400$-$5000$. We show how one can recover 2D and tomographic power spectra from the full 3D cosmic shear power 
spectra and present a measurement of the 2D cosmic shear power spectrum, and measurements of a set of $2$-bin and $6$-bin 
cosmic shear tomographic power spectra; in doing so we find that using the 3D power in the calculation of such 2D and 
tomographic power spectra from data naturally accounts for a minimum scale in the matter power spectrum. We use 
3D cosmic shear to constrain cosmologies with parameters $\Omega_{\rm M}$, $\Omega_{\rm B}$, $\sigma_8$, $h$, $n_s$, $w_0$, $w_a$. 
For a non-evolving dark energy equation of state, and assuming a flat cosmology, lensing combined with WMAP7 results 
in $h=0.78\pm 0.12$, $\Omega_{\rm M}=0.252\pm 0.079$, $\sigma_8=0.88\pm 0.23$ and $w=-1.16\pm 0.38$ using only scales 
$k\leq 1.5h$ Mpc$^{-1}$. We also present results of lensing combined with first year Planck results, 
where we find no tension with the results from this analysis, but we also find no significant improvement over the 
Planck results alone. We find evidence of a suppression of power compared to LCDM on small scales $1.5 < k\leq 5.0h$ Mpc$^{-1}$ 
in the lensing data, which is consistent with predictions of the effect of baryonic feedback on the matter power spectrum. 
\end{abstract}

\begin{keywords}
Cosmology: cosmological parameters. Gravitational lensing: weak
\vspace{-0.9cm}
\end{keywords}
\newpage
\section{Introduction}
Light from distant galaxies is gravitationally lensed as a result of 
mass perturbations along the line of sight. In the weak-field regime, away from the critical curve of the lensing mass, 
the effect is to change the observed projected ellipticity
of light bundles, called shear, caused by the
tidal field generated by the intervening mass; so-called weak
lensing. In our Universe weak lensing of light from distant galaxies is caused
by the distribution of matter in large-scale structure, an effect
called cosmic shear. Because our view of the Universe is inescapably
three dimensional -- we observe galaxies across the sky, but they are also
spread in distance, or redshift -- what we observe 
is characterised by a three dimensional cosmic shear field. The
use of both the shear information and the full redshift information is a
technique called 3D cosmic shear, and it is the focus of this paper. 

3D cosmic shear was first presented in Heavens (2003), where it was
suggested that it may be a particularly sensitive probe of dark
energy. The methodology was further developed in Castro, Heavens,
Kitching (2003); Kitching, Heavens, Miller (2011) and Munshi et al. (2011). 
The method works by 
representing the three dimensional cosmic shear field using spin-2 spherical
harmonics and spherical Bessel functions, where the signal is the set of 
coefficients calculated as a sum over the measured shears for a population of
galaxies. Fisher matrix predictions for wide
field imaging surveys were made in Heavens et al. (2007) and Kitching (2007),
where it was shown that 3D cosmic shear is a sensitive
probe of the dark energy equation of state because it is a function of
both the geometry of the Universe and of the growth of structure.
In addition to dark energy properties it has been shown that 3D
cosmic shear can measure minimally modified
gravity parameters (Heavens, Kitching, Verde 2007), the total sum of
neutrino mass (Kitching et al., 2008) and possibly even the neutrino hierarchy 
(de Bernardis et al., 2009; Jimenez et al., 2010). 
3D cosmic shear was applied to data as a proof of concept in Kitching
et al. (2007) on the COMBO-17 survey that covered approximately $1.5$
square degrees, and presented a conditional error on a constant dark energy
equation of state $w$ in line with Fisher matrix predictions.

3D cosmic shear is a method that works in spherical Bessel/spherical harmonic space
(the spherical coordinate analogue of Fourier space, both being eigenfunctions of the 
Laplacian operator), and does not bin information in the
redshift direction. There are several approximations to 3D cosmic
shear that have been used or proposed, the most widely cited being 2D
correlation function analyses and 2D cosmic shear power spectra. The
generalisation from 2D correlation functions or power spectra 
to a series of projected 2D slices in redshift is referred to as
so-called `tomography' (e.g. Hu, 1999), where intra-bin correlations are supplemented with
inter-bin correlations\footnote{The word tomography, meaning a cross-section, slice or 
image of a predetermined plane in the body, is used colloquially in weak lensing to refer to 
the power spectra of projected planes integrated along the line of sight.}.
Each of these is related to 3D cosmic shear 
by various steps and approximations namely 1) the Limber approximation 
(e.g. LoVerde \& Afshordi, 2008),
2) a transform from radial spherical harmonics to 2D Fourier space
(Fourier in angle but real-space in redshift direction),
on each tomographic slice, 3) a binning in redshift (Kitching,
Heavens, Miller, 2011) and possibly 4) a
further Fourier transform from Fourier space to real-space in
angle. Both the Limber approximation and redshift space binning 
result in a loss of information, whereas the spherical harmonic and Fourier transforms are in
principle lossless, but in practice cause the relationship between radial and
angular scales to become more involved. As a result 3D cosmic shear has
several features that make it a useful technique:
\begin{enumerate}
\item 
It does not bin the data in redshift, but uses every galaxy
individually. This has the advantage that information is not lost,
particularly along the direction (redshift) in which discoveries about
dark energy are likely to appear -- dark energy affects the rate of change of
the expansion history of the Universe. This is in contrast to 2D and
tomographic methods that bin and average in redshift thereby losing
information.  
\item
It allows for a control of scales included in the analysis in a
rigorous manner, both angular ($\ell$) and radial ($k$) modes can be treated
independently. As a result problematic regions, at small scales, for
example due to inaccuracy in the modelling of the non-linear growth of structure
(e.g. Smith et al., 2003; Takahasi et al., 2012) 
or baryon feedback (e.g. van Daalen et al., 2011; Semboloni et al., 2011, 2013) 
in the dark matter density field, can be down-weighted. This is in contrast to real-space correlation function
techniques where scales are less easy to disentangle, and 2D power spectrum methods where
radial modes are necessarily linked to angular modes through the
Limber approximation.
\item
Because each individual galaxy is used in the estimator, rather than
averaged quantities, uncertainties on individual galaxy measurements 
can be used explicitly. In a Bayesian approach this means including
posterior probability information on measured quantities; for example
the photometric redshift probability $p_g(z)$ for each galaxy (as shown in Kitching et
al., 2011) but also potentially posterior information on galaxy shapes, or 
surface brightness distributions.  
\item 
The formalism uses a one-point estimator as the signal data vector with the 
cosmological sensitivity encoded in the covariance (the mean is zero). The one-point 
estimator encodes the full 3D field. The covariance is calculated 
analytically and therefore does not need to be estimated \emph{ad hoc} from 
the data or simulations, which avoids issues of convergence and limitations due 
to the finite number of simulations (see Taylor, Joachimi, Kitching, 2013); 
although this benefit is eroded somewhat 
by the assumption in this paper of a Gaussian likelihood function for the transform 
coefficients.  
\end{enumerate}
In this paper we apply 3D cosmic shear for the first time to a
wide-field weak lensing data set, CFHTLenS (Heymans et al., 2012;
Erben et al., 2013). The CFHTLenS survey covers $154$ square degrees
and uses state-of-the-art weak lensing measurements (\emph{lens}fit, Miller et al., 2013) and
photometric redshift measurements ({\sc BPZ}, Hildebrandt et al., 2012), in addition the 
combined weak lensing and redshift measurements are the first to be rigorously tested for systematics (Heymans et al., 2012).  
When accounting for masking and systematics $61$ per cent of the data ($171$ pointings in total) 
has been shown to be fit for purposes of cosmic shear science (Heymans et al. 2012). 

This analysis is not a proof
of concept but a demonstration that 3D cosmic shear can constrain cosmological 
parameters to a level comparable to other currently available
cosmological probes even over a relatively small area survey. In
addition we extend the methodology to account for survey masks using a
3D pseudo-$C_{\ell}$ methodology, using the formalism which was first presented in Munshi et
al. (2011). 3D cosmic shear techniques have also been investigated by Ayaita et
al. (2012), and 3D `Fourier-Bessel' approximations to spherical
harmonic transforms for spin-0 fields have been presented in Leistedt
et al. (2012).

This analysis\footnote{The software used in this paper, {\sc 3DFast}, that
includes the 3D cosmic shear estimators and parameter estimation, is
available on request, and more details are available here \url{http://www.thomaskitching.net}.} 
is independent of, and conservative in respect to, the cosmological analysis of
CFHTLenS from the 2D and tomographic correlation function results presented in 
Kilbinger et al. (2013), Simpson et al. (2013), Benjamin
et al. (2013) and Heymans et al. (2013), all of which were based on the
same software ({\sc athena}\footnote{\url{http://www2.iap.fr/users/kilbinge/athena}} and {\sc
  nicaea}\footnote{\url{http://www2.iap.fr/users/kilbinge/nicaea}}) and the same 
simulations (Harnois-Deraps et al., 2012) where each analysis varied the output 
parameter set. Heymans et al. (2013) present a coarsely-binned correlation
function measurement, $6$ bins in redshift, that includes the additional estimation of a
parameter that encodes intrinsic alignment (IA) systematics 
(Hirata \& Seljak, 2004). In this
paper we address intrinsic alignments by explicitly removing photometrically identified 
early-type galaxies from the analysis, which have a non-zero IA signal in Heymans et
al. (2013). Similar motivation is found in Mandelbaum et al. (2011) who found a null IA signal using 
the WiggleZ dark energy survey, 
that had a galaxy sample that was comparable in galaxy type and redshift selection as the late-type galaxies in CFHTLenS.  

This paper is presented as follows. In Section \ref{Photometric 3D
  Shear Estimator} we summarise the 3D cosmic shear method, in Section
\ref{Data} we present some approximations to the data including measurements of 2D
and tomographic power spectra, in Section \ref{Results} we present the
cosmological parameter constraints. Conclusions are drawn in Section
\ref{Conclusion}. Mathematical details are presented in a series of 
Appendices. 

\section{Methodology}
\label{Photometric 3D Shear Estimator}

In a 3D cosmic shear likelihood analysis the data vectors are a set of spherical harmonic transform 
coefficients, and it is the covariance of these coefficients that contains cosmological information. 
Here we describe the data vectors, covariance and the likelihood function. 

\subsection{The data vectors}
\label{The data vectors}
3D cosmic shear expresses the three dimensional shear field in terms of its spherical Bessel/spherical harmonic coefficients
(Heavens, 2003; Castro, Heavens, Kitching, 2003; the CFHTLenS fields are small enough to use the flat sky exponential approximation)  
\be 
\label{a}
\gamma_i(k,\ell)=\sqrt{\frac{2}{\pi}}\sum_{g}e_{g,i} j_{\ell}(kr^0_g){\rm e}^{-\idot\bell.\btheta_g}W(r^0_g),
\ee
where $k$ are radial wavenumbers and $\ell$ are angular wavenumbers; $\bell$ is a 2D wavenumber on the sky 
where $\bell=\ell_x+\idot\ell_y$ and $\ell=\sqrt{\ell_x^2+\ell_y^2}$.
Equation (\ref{a}) is a sum over galaxies, 
weighted by a spherical Bessel function $j_{\ell}(kr)$, exponential terms, and an arbitrary weight function $W$. 
$e_{g,i}$ are the $i^{\rm th}$ components of ellipticity, $i=\{1,2\}$, for galaxy $g$ at a 3D angular 
and radial coordinate ($\btheta_g$, $r^0_g$); we use $e$ for the observed quantity and $\gamma$ as the computed quantity. One can also use a facultative 
factor of $k$ in the transform (as used in Castro, Heavens, Kitching, 2003) but results are unchanged. 
This is a one-point estimator describing a 3D shear field. 
 
As explained in Kitching, Heavens, Miller (2011) $r^0_g$ is a distance, not a redshift, and so requires the assumption of a fixed reference cosmology; 
this assumption is benign since the $j_{\ell}(kr)$ simply acts as a weight for both the data and theory. In this paper the 
distance $r^0_g$ is estimated from the 
maximum posterior photometric redshift for each galaxy. 
This expression assumes a flat sky approximation (replacement of $Y_{\ell}^m$ functions with complex exponentials) but this can 
in principle be relaxed (see Castro, Heavens, Kitching, 2003). This expression also technically assumes a flat Universe 
through the use of the 
spherical Bessel functions but again this can be relaxed resulting in the use of hyperspherical Bessel functions; 
however the hyperspherical Bessel function is very close to the spherical Bessel function (see for example Kosowsky, 1998), and 
in any case post-WMAP  
our cosmological model is observed to have only small perturbations about flatness, if at all (e.g. Hinshaw et al., 2013 
constrain $\Omega_{\rm K}=-0.0027 [+0.0039/-0.0038]$).
Note that the Limber approximation is not equivalent to a flat sky approximation: the Limber approximation 
links $k$ and $\ell$-modes by effectively replacing spherical Bessel functions with delta functions (LoVerde \& Afshordi, 2008) and is not used here; the 
flat sky approximation replaces spherical harmonics $Y_{\ell}^m$ with exponentials (Castro, Heavens, Kitching, 2003). 

Equation (\ref{a}) is calculable from the data, given a set of ellipticity estimates and results in four data vectors which come from the 
real and imaginary parts of the ellipticity and exponential terms. In Kitching et al. (2007) these four data vectors were all used in the 
likelihood calculation, however these terms can be separated into two $E$-mode data vectors and two $B$-mode data vectors as shown in Appendix A, resulting in 
real and imaginary data vectors: ${\mathbb R}[\gamma_E(k,\ell)]$, ${\mathbb I}[\gamma_E(k,\ell)]$, ${\mathbb R}[\gamma_B(k,\ell)]$, ${\mathbb I}[\gamma_B(k,\ell)]$. 
In the cosmological analysis the $E$-mode is expected to contain the signal, whereas the $B$-mode should be consistent with shot noise.

For the CFHTLenS analysis (described in Miller et al., 2013; Heymans et al. 2012, Erben et al., 2013) there are three changes 
that must be applied to the catalogues in order to create unbiased estimators of the transform coefficients:  
\begin{enumerate}
\item
weighting by the shape measurement (\emph{lens}fit) weight $W(r^0_g)=W_{L,g}$,
\item
application of the $e_2$ additive calibration correction $c_{2,g}$,
\item
application of the multiplicative ellipticity $(1+m_g)$ calibration correction,
\end{enumerate}
all of which vary for each galaxy $g$. 
The \emph{lens}fit weight is an inverse-variance weight that 
encapsulates the confidence in the ellipticity measurement (galaxies measured with a sharply peaked likelihood in ellipticity have a higher weight), 
as well as population variance of the ellipticity estimates on a galaxy-by-galaxy basis (see Miller et al., 2013 section 3.6).
The calibration corrections relate the observed ellipticity $e_{\rm obs}$ to an estimate of the true ellipticity $e_g$ for each galaxy 
through a linear relation $e_g=(1+m_g)e_{\rm obs}+c_g$ for each ellipticity 
component. The $c$ term is an empirical bias correction applied to the CFHTLenS catalogue under the assumption that 
the expected value is zero $\langle c\rangle=0$: the $c_1$ component is consistent with zero but the $c_2$ term is non-zero.   
The multiplicative term is signal-to-noise and galaxy-size dependent and is calibrated with respect to image simulations of CFHTLenS; Heymans et al. (2012) 
provide an empirically fitted formula to simulations to account for this bias. 
For the calculation of the spherical harmonic coefficients we first subtract the additive $c_2$ component from each galaxy ellipticity, we then modify the 
spherical harmonic coefficients as described in Appendix B where we show that the multiplicative term results in a scaling of the coefficients and also a 
mixing of the E and B-modes that must be accounted for. In calculating the coefficients we sum over all galaxies defined in Section \ref{Data}. 

A further modification to the data vectors is that the angular coordinates 
on the sky ($\alpha$, $\delta$) (Right Ascension and declination) need 
to be converted into tangent-plane (flat sky) coordinates ($\theta_x$, $\theta_y$); this is 
achieved using spherical trigonometry using a gnomonic projection where $\cos\theta_x=\cos^2(\pi/2-\delta)+\sin^2(\delta-\pi/2)\cos\alpha$ 
and $\theta_y=\delta$, and angles are converted; for each field we use the mean of the coordinates as the central coordinates 
for the projection. 
This projection is not a limitation of the methodology, but is needed when making the flat-sky 
assumption in this paper.

\subsection{Covariance}

\subsubsection{Signal Covariance}
For 3D cosmic shear the transform coefficients have an expectation value of zero but the expected covariance is non-zero 
and it is that which 
is used as the cosmology-dependent signal (see Section \ref{Likelihood}). As described in Kitching, Heavens, Miller (2011) the 
lensing part of the covariance, which assumes only the cosmological principle, but not Gaussianity, can be written
\ba
\label{cov}
C^S_{\ell}(k_1,k_2)&=&(D_1^2+{\rm i}D_2^2)\frac{4}{\pi^2c^2}A^2\sum_g\sum_h
\left[j_{\ell}(k_1 r^g_0)j_{\ell}(k_2 r^h_0)\right]\nn
&&\int{\rm d}z' p_g(z') \int{\rm d}z'' p_h(z'') \nn
&&\int_0^{r(z')}{\rm d}{\tilde r}\int_0^{r(z'')}{\rm d}{\tilde{\tilde r}}
F_K(r',{\tilde r})F_K(r'',{\tilde{\tilde r}})\nn
&&\int \frac{{\rm
    d}k'}{k'^2}\frac{1}{a({\tilde r})a({\tilde{\tilde r}})}
j_{\ell}(k'{\tilde r})j_{\ell}(k'{\tilde{\tilde
    r}})\sqrt{P(k';{\tilde r})P(k';{\tilde{\tilde r}})},\nn
&&
\ea
where $p_g(z)$ and $p_h(z)$ are the posterior probabilities that galaxies $g$ and $h$ are at redshift
$z$. $P(k;r)$ is the matter power spectrum; the functions $F_K(r,r')=S_K(r-r')/[S_K(r)S_K(r')]$ where $S_K(r)=\sinh(r)$, $r$, $\sin(r)$ for cosmologies with 
spatial curvatures $K=-1$, $0$, $1$; $a(r)$ is the dimensionless scale factor. The pre-factor 
$A=3\Omega_{\rm M} H_0^2/2$ where $H_0$ is the current value of the Hubble parameter and $\Omega_{\rm M}$ is the ratio of the total matter density to the 
critical density. Where we label with a semicolon e.g. $P(k; r)$, the comoving distance is labelling the 
time-dependence $P(k,r[t])$ (Castro, Heavens, Kitching, 2003). The sums are over 
galaxies used to construct the data vectors. 
This is a slightly more general expression than that in Kitching, Heavens, Miller (2011) where here we explicitly 
include the Fourier derivatives $D_1=\frac{1}{2}(\ell_y^2-\ell_x^2)$ and
$D_2=-\ell_x\ell_y$ that convert from potential to shear (see Appendix A). The covariance then has real and imaginary parts which are not necessarily 
equal for each $(\ell_x,\ell_y)$ mode; in the analysis we treate the real and imaginary parts separately. 

The photometric redshift uncertainties for each galaxy enters the covariance calculation 
as shown in equation (\ref{cov}). Note that the photometric uncertainty does not enter into the data vector where 
the maximum likelihood redshift for each galaxy is used; the covariance then accounts for the scatter 
in the data vector caused by this assumption as discussed in Kitching, Heavens, Miller (2011).

This is the signal part of the covariance of the $\gamma_E(k,\ell)$ where 
$C^S_{\ell}(k_1, k_2)=\langle {\mathbb R}[\gamma_E(k_1, \ell)]{\mathbb R}[\gamma_E(k_2, \ell)]\rangle+{\rm i}\langle {\mathbb I}[\gamma_E(k_1, \ell)]{\mathbb I}[\gamma_E(k_2, \ell)]\rangle$. 
Throughout this investigation we use {\sc camb}\footnote{\url{http://camb.info} version October 2012} to calculate the matter power spectra with the {\sc halofit} 
(Smith et al., 2003) non-linear 
correction and the module for Parameterized Post-Friedmann (PPF) prescription for the dark energy 
perturbations (Hu \& Sawicki, 2007; Fang et al., 2008; Fang, Hu \& Lewis, 2008). 

\subsubsection{Noise Covariance}
\label{Noise Covariance}
The shot noise part of the covariance is given by Kitching et al. (2007) as   
\be 
\label{noi}
N_{\ell}(k_1,k_2)=\frac{\sigma^2_{\epsilon}\Delta\Omega}{4\pi^2}\int {\rm d}z \bar n(z) j_{\ell}(k_1 r^0)j_{\ell}(k_2 r^0),
\ee
where $\bar n(z)=\sum_g p_g(z)$ is the sum of the posterior redshift probabilities, and $\sigma^2_{\epsilon}$ 
is related to the variance of the ellipticity distribution in the data. 
The E and B-mode separation involved in manipulating the spherical harmonic coefficients for use in the likelihood evaluation 
(see Appendix A and B) causes the variance $\sigma^2_{\epsilon}\in {\mathbb C}$ to be related to variance of the observed ellipticities as described in Appendix B. 
$\Delta\Omega$ is the solid angle, or area, of the survey. 

The shot noise is calculated assuming a reference 
cosmology, for which we use the best-fit WMAP7 values (Komatsu et al., 2011); this is a benign choice 
as long as the same reference cosmology is used in the data vector calculations. We do not consider any cross-terms between the noise and signal, which are expected to be zero in 
the absence of source-lens clustering (see e.g. Valageas, 2013). 

\subsubsection{Pseudo-$C_{\ell}$}
\begin{figure}
\psfig{file=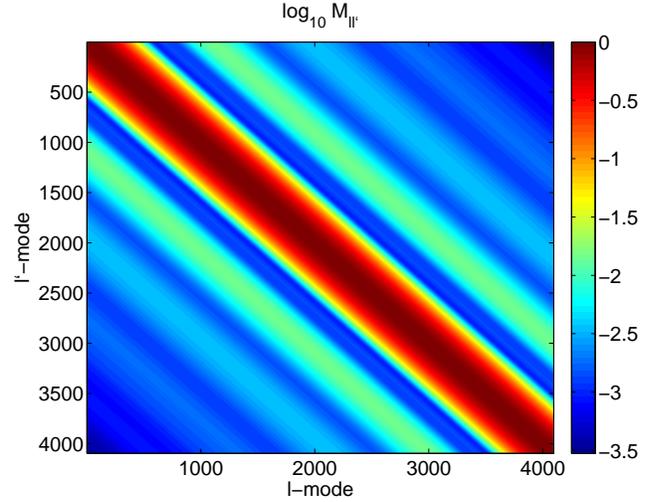, width=\columnwidth, angle=0}
\caption{Normalised 3D mixing matrix for the W1 field, the colour-scale shows the amplitude of the mixing matrix, and is logarithmic 
as depicted.}
\label{fmll}
\end{figure}

A further sophistication applied here is the accounting for angular masks in the data. To account for masks we adapt the pseudo-$C_{\ell}$ methodology 
(that has been used in CMB studies, e.g. Hivon, 2002) for use with  
3D spherical harmonics in Appendix C. Masks in data act to move power from the angular scale of the masks to other parts of the power spectrum, this `mixing' of 
power can be calculated given the mask and expressed as a mixing matrix $M^{3D}_{\ell\ell'}$. Here we multiply the theoretical 3D power by the mixing matrix to simulate the effect of the mask 
and compare this with the data; this results in equations (\ref{pcoeff}) to (\ref{c}), the latter of which is reproduced here
\be 
\widetilde C^{EE}_{\ell}(k_1,k_2)=\left(\frac{\pi}{2}\right)^2\sum_{\ell'}\left(\frac{\ell'}{\ell}\right)M^{3D}_{\ell\ell'}
C^S_{\ell}\left(k_1\frac{\ell'}{\ell},k_2\frac{\ell'}{\ell}\right),
\ee
where $\widetilde C^{EE}_{\ell}(k_1,k_2)$ is the pseudo-$C_{\ell}$ estimator of the 3D power and $M^{3D}_{\ell\ell'}$ is the 3D mixing matrix defined in equation (\ref{c}); this is the expression we use to account for the masks, on the theory
side, in the likelihood analysis presented. Alternatively one could attempt to invert the mixing matrix and apply this to the data to undo the effect of the masks (although in this case regularisation of the matrix may be required, for example binning in $\ell$-mode, 
depending on its condition number). The mixing matrix only affects the signal part of the covariance matrix; the effect of the mask on the shot noise covariance is a simple area scaling.  
We show the mixing 
matrix calculated for the W1 field (see Section \ref{Data} for a description of the data) in Figure \ref{fmll}. This formulation, presented in Munshi et al. (2011), is a generalisation of a 2D pseudo-$C_{\ell}$
formalism (e.g. Hikage et al., 2011; Kitching et al., 2012 Appendix A) to include a correct weighting over all radial $k$-modes that contribute to each convolved $\ell$-mode, the application of a 2D mixing matrix in this case would
not be sufficient for a 3D cosmic shear analysis. The mixing matrix is applied to the real and the imaginary parts of the signal covariance in the same way, although in principle there could exist systematics errors that create mixing matrices that do not have this property (see Kitching et al. 2012). 

In Kitching et al. (2007) a correction was made to the covariance matrix to account for the very small angular size of the COMBO-17 field. 
In this paper we do not apply this correction because the survey geometry of the CFHTLenS fields is large enough that the correction factor (${\mathcal F}$, equation 10 in Kitching et al., 2007) is approximated by a delta function 
and also that the mixing matrix formalism itself consistently accounts for the survey geometry.

We can now define the observed power spectrum as the sum of the pseudo-$C_{\ell}$ signal and the noise matrix 
\be 
C_{\ell}(k_1,k_2)=\widetilde C^{EE}_{\ell}(k_1,k_2)+N_{\ell}(k_1,k_2).
\ee
which we refer to as the 3D cosmic shear power spectrum. Recall that these are complex valued power spectra.

These covariance estimates represent a significant computational task: involving 2 nested sums over the galaxy 
population, 5 nested integrals, computation of the matter power spectra, and 
the matrix sum with the mixing matrix. Previous implementations (Kitching et al., 2007) were prohibitively 
slow ($\sim$ 1 hour per cosmology on a desktop computer in 2007), 
limitations that have been overcome in the {\sc 3DFast} implementation 
($\sim$ 5 seconds per cosmology on a desktop computer in 2013; with a parallelised and extendable code) 
allowing for exploration of large cosmological parameter sets.

\subsection{Likelihood}
\label{Likelihood}
The likelihood of a complex random field in Fourier or spherical harmonic space is more involved than simply treating the real and imaginary parts of the field 
independently. Here we first describe the covariance matrix for 3D cosmic shear and then define the likelihood function. 

\subsubsection{The Affix-Covariance}
A sophistication that we apply for the first time in a cosmological context here, is modification to the likelihood function that the complex nature
of the field requires (see Alsing et al. in prep; for more details). 
As shown by Picinbono (1996) and Nesser \& Massey (1993) for a normally distributed complex quantity $z=x+{\rm i}y$, 
where $x\in {\mathbb R}$ and $y\in {\mathbb R}$, the joint probability distribution for the two quantities must be written
\be
p(x,y)=p(z,z^*)=\frac{1}{\pi^2|A|^{1/2}}{\rm exp}\left(- \frac{1}{2}Z^{\dagger} A^{-1} Z\right)
\ee
where $Z=(z,z^*)^T$, and $\dagger$ refers to a Hermitian-conjugate. 
We refer to the matrix $A$ as the affix-covariance\footnote{Affix is a word that can refer to a complex number (Whittaker \& Watson, 1990), but it also means `fasten' or `attach'.} (one may also refer to this as a pseudo-covariance, but we wish to avoid confusion with the transformation 
required to account for a survey mask). $A$ contains two sub-matrices: the usual covariance matrix $\Gamma = \langle zz^{*T}\rangle$ and the so-called relation matrix $R = \langle zz^T\rangle$. 
In general, the covariance matrix alone is not sufficient to fully specify the second-order statistics of a complex random variable, and 
care must be taken when transforming to harmonic space so that all possible correlations are retained. In particular, the off-diagonal blocks in the affix-covariance which vanish 
for a real-space variable may no longer vanish in harmonic space; this is a common pitfall when dealing with complex random variables in Fourier or harmonic space. 
The case when the relation matrix vanishes is a condition known as second-order circularity, but this does not apply here. 

In the case of 3D cosmic shear we have a data vector that consists of the real and imaginary coefficients
$\gamma_E(k,\ell)={\mathbb R}[\gamma_E(k,\ell)]+{\rm i}{\mathbb I}[\gamma_E(k,\ell)]$ such that we can define the affix-covariance matrix, for 
each $\ell$-mode, as a $2N_k\times 2N_k$ matrix,
where $N_k$ is the number of $k$-modes in the coefficients 
\be
A_{\ell}(k_1,k_2)=\left( \begin{array}{cc}
\Gamma &  R\\
R^T & \Gamma^* \\
\end{array} \right)_{\ell}
\ee
which consists of four $N_k\times N_k$ blocks that relate to the 3D cosmic shear power spectra
\ba
\Gamma_{\ell}(k_1,k_2)&=&{\mathbb R}[C_{\ell}(k_1,k_2)]+{\mathbb I}[C_{\ell}(k_1,k_2)]\nn
R_{\ell}(k_1,k_2)&=&{\mathbb R}[C_{\ell}(k_1,k_2)]-{\mathbb I}[C_{\ell}(k_1,k_2)]
\ea
note that the relation matrix $R$ will not depend on the shot noise for some $\ell$-modes as the two contributions will cancel.  

\subsubsection{Likelihood Function}
Given the data vectors and the theoretical covariance, the cosmological parameter likelihood function (assumed Gaussian) can now be written as 
\ba
\label{el}
L(p)=\sum_{\ell}\frac{1}{\pi^2 |A_{\ell}|^{1/2}}{\rm exp}\left[-\frac{1}{2}\sum_{kk'}Z_{\ell}(k)A^{-1}_{\ell}(k,k')Z_{\ell}(k')\right]
\ea
where $Z_{\ell}(k)=(\gamma_E[k,\ell],\gamma^*_E[k+N_k,\ell])^T$. The sums over $\ell$ and $k$ are over the scales defined in Section \ref{Scales}. 
We label the parameters of interest $p$. 
 
In the 3D cosmic shear formalism, which uses a one-point data vector of spherical harmonic coefficients, 
the covariance itself contains all the cosmological information, and the inverse is exact. Hence there is no need to estimate the
covariance itself from data or simulations. When
the covariance must be estimated this results (because of the Wishart-distribution of the covariance) in the need for calibration with simulations
(see for example Taylor, Joachimi, Kitching, 2013) to account for the Kaufman/Anderson bias (Kaufman, 1967; Anderson, 2003; see also Hartlap 
et al., 2007). For the case of correlation function analyses of the CFHTLenS data Kilbinger et al., (2013)
used a hybrid ansatz of a combined analytical and estimated covariance, the former does not need to be corrected for the Kaufman/Anderson bias but the latter does. As discussed in Taylor, Joachimi, Kitching (2013) one mitigation approach is to use an analytic covariance, as is done in this paper. 

\begin{figure*}
\psfig{file=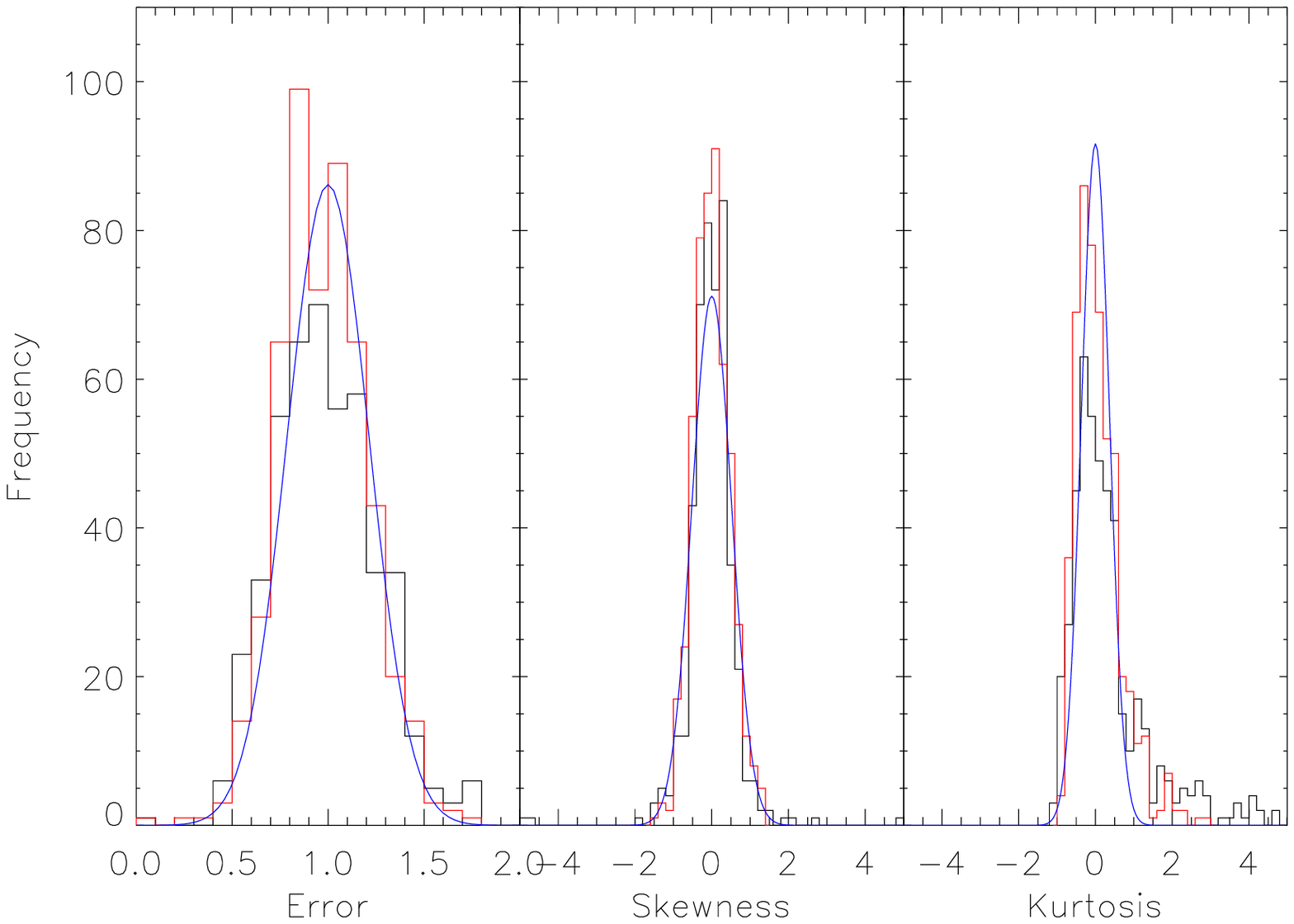, width=\columnwidth, angle=0}
\psfig{file=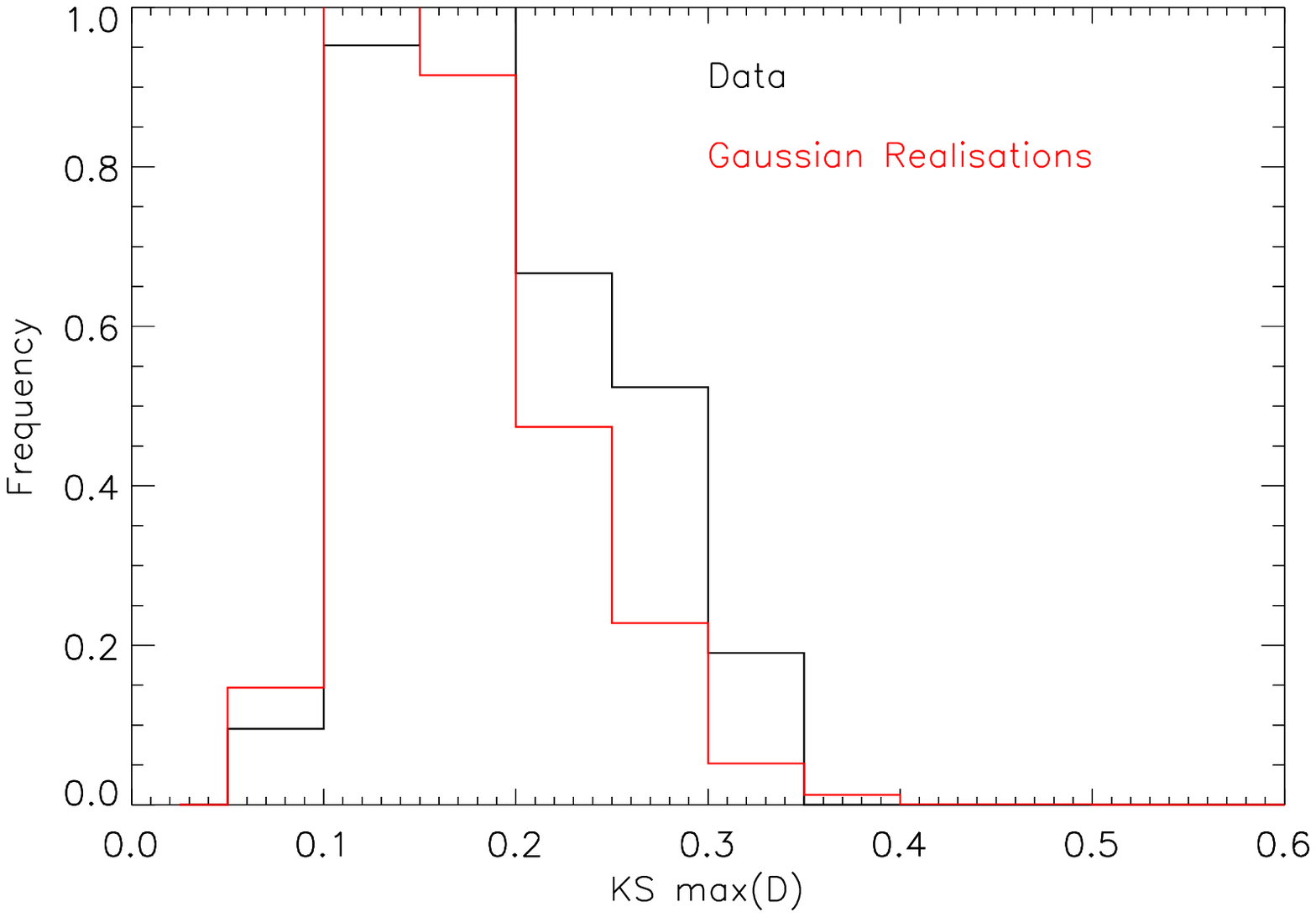, width=\columnwidth, angle=0}
\caption{Left: The distribution of standard deviations, skewness and kurtosis 
for the normalised transform coefficients over all $\ell$-modes and for all CFHTLenS fields. We show the distribution of the data (black), 
the distributions from a Gaussian realisation of the data (red) and the analytic expected distribution of these statistics (solid blue lines) 
for a data set of this size (see Section \ref{Data}).  
Right: A normalised histogram of Kolmogorov-Smirnov (KS) supremum values, over all $\ell$-modes from all CFHTLenS fields, 
between the distribution of the normalised transform 
coefficients and a unit Gaussian 
distribution (black). We compare this with a similar set of KS supremum values between a unit Gaussian and Gaussian realisations of the data (red), 
where any difference is due to noise only. These KS supremum value distributions are consistent 
(with mean values and errors of $0.17\pm 0.06$ and $0.21\pm 0.11$).}
\label{hist}
\end{figure*}

An assumption we make here, that the likelihood function is Gaussian, 
is likely to be incorrect in detail on small scales, but for CFHTLenS this approximation is
sufficient. For a survey the size of CFHTLenS we expect to be in the shot noise-dominated regime for any individual $\ell$-mode, and one may expect 
that the shear coefficients will be Gaussian distributed because of the central limit theorem acting through the sum that is performed over the 
real-space galaxy shear values (that may have a non-Gaussian distribution). As a test of the 
Gaussianity of the shear coefficients we examine their distribution divided by the expected shot noise (equation \ref{noi}), what we will refer to as 
`normalised transform coefficients'; which should have a unit Gaussian distribution. 
We show in Figure \ref{hist} a histogram of the variances, skewnesses and kurtoses of the normalised transform coefficients over all $\ell$-modes over  
all CFHTLenS fields (see Section \ref{Data} for a description of the data); we compare this to the expected distribution of these statistics for Gaussian 
distributed data of this size (for the expected error on the error see Taylor, Joachimi, Kitching, 2013; for the skewnesses see Kendall et al., 1998; 
for the kurtoses see Keeney \& Keeping, 1962), and also to a mock realisation of the normalised transform coefficients sampled from a unit Gaussian. 
We find that there is no significant deviation from Gaussianity. We do find that $\ls 1\%$ of the modes have a small positive excess kurtosis, but 
due to the overall weak constraints presented in Section \ref{Results} this level of non-Gaussianity should not impact results. 
In Figure \ref{3dcs} we also show histograms of the normalised transform coefficients for 
a representative set of $\ell$ values, and averaged over all $\ell$ values for information. 
As a quantitative test of Gaussianity we also perform a Kolmogorov-Smirnov (KS) comparison between the distribution of the 
transform coefficients and the best-fit Gaussian, and compare this with Gaussian realisations of the data using the best-fit values. 
The results of this KS comparison are shown in Figure \ref{hist}, where we again find no significant deviation from Gaussianity, 
although the KS test is best at detecting a shift in the mean of two distributions. 

For an isotropic Gaussian field, the harmonic coefficients are statistically independent and normally distributed, where 
the magnitude of each coefficient provides an independent estimate for the power at the scale of the coefficient (this is where 
the cosmological constraints stem from in the analysis used in this paper).  Isotropy and the high shot noise ensure that the 
coefficients are close to Gaussian distributed, they have an enhanced variance because of non-linear growth of the power spectrum, 
and are correlated for different angular modes only through the angular window. All these effects are all included in the analysis, 
but we make an approximation, in assuming a multivariate Gaussian likelihood, that combinations of coefficients which are 
uncorrelated are also taken to be statistically independent.

We assume in the likelihood calculation that the data vector of shear coefficients is Gaussian distributed, we do not 
assume Gaussianity of the shear field itself, as stated before equation (\ref{cov}).
Since we use a transform of the field itself, rather than power spectra or correlation functions,
the covariance of the data is a $2$-point quantity rather than $4$-point. We only require isotropy (not Gaussianity) to
have a diagonal covariance matrix for the matter power spectrum, and we include nonlinearities by using the nonlinear
power spectrum. With 2-point statistics, the assumption of gaussianity underestimates the (4-point) covariance,
but here this is not the case.
Therefore it is sufficient to show that the shear transform coefficients have a Gaussian distribution, and
the analysis shows that this is the case, with only very small departures from Gaussianity in the coefficients.
The analysis uses the covariance of the shear field as the cosmologically sensitive statistic, hence if the survey area was
small this could be sensitive to sample variance whereby
local fluctuations could result in an inferred cosmology significantly different from a global description.
However because the CFHTLenS survey is a large area, much large than $1$ square degree, such
sample variance effects are expected to be minimal (e.g. Driver \& Robotham, 2010).

\subsection{Tests on simulations}
Despite the fact that the covariance does not need to be estimated from simulations, we nevertheless wished to test the 
formalism and code to confirm that it was performing as expected. To do this we used the {\sc SUNGLASS} simulations from 
Kiessling et al. (2011), which contain shear and redshift probability information assigned on a galaxy-by-galaxy basis, which 
are ideal for the testing of 3D cosmic shear in principle. 
For 3D cosmic shear we could not use the CFHTLenS {\sc Clone}
(Harnois-Deraps et al., 2012; used in Kilbinger et al., 2013, Simpson et al., 2013, Benjamin
et al., 2013, Heymans et al., 2013 for calibration of the covariance)
because the shears were computed in a series of discrete redshift slices not from a full three
dimensional shear field. 
The {\sc SUNGLASS} simulations used here 
consist of $150$ realisations of $100$ square 
degrees with a galaxy number density of $16$ galaxies per square arcminute with a median redshift of $0.75$, 
and are therefore well-matched to CFHTLenS (Heymans et al., 2012) survey characteristics in terms of number density and 
redshift distribution. The simulations however are smaller in terms of area and do not include an 
intrinsic alignment model or survey masks, but these are not insurmountable issues and should be addressed in 
subsequent simulations. 

However a more serious limitation is that in analysing these simulations we had to use a limited $k$ and $\ell$ range.  
In the radial direction we set a limit\footnote{We will use the same limit for the data analysis, but this is a coincidence.} of 
$k\leq 1.5 h$Mpc$^{-1}$, by referring to Figure 3 in Kiessling et al. (2011) where it can be seen that the 
predicted tomographic $C(\ell)$ begins to deviate from the simulated power at $\ell\gs 1.5 r(z_{\rm bin})$ where 
$r(z_{\rm bin})$ is the comoving distance of the tomographic bin. Also in Kiessling et al. (2011) 
a conservative cut in $\ell$ was used of $500\leq \ell\leq 1000$; 
this is the regime where the box-size (on the large scales) and particle 
resolution (on the small scales) do not affect the 
fidelity. The limitations mean that the results of testing on these simulations are expected to give much larger error bars than one should 
get from data -- where larger $k$ and $\ell$ ranges may be used -- which is not limited by simulation resolution effects. 

We calculated the likelihood surfaces in the ($\sigma_8$, $\Omega_{\rm M}$) plane over the $150$ simulations, 
over ranges of $0.1\leq \sigma_8 \leq 3.0$ and $0.1\leq \Omega_{\rm M}\leq 0.9$, 
using the likelihood function described in Section \ref{Likelihood}; all other 
cosmological parameters were fixed at the values provided in Kiessling et al. (2011), where the simulations used $\sigma_8=0.8$, $\Omega_{\rm M} = 0.27$,
$\Omega_{\rm B} = 0.045$, $n_s = 0.96$, and $h = 0.71$. We find the mean maximum likelihood values in this test are 
$\Omega_{\rm M}=0.27\pm 0.020$ and $\sigma_8=0.82\pm 0.056$, which are consistent with the input cosmology. 
Within the limitations of the simulations available 
we find that the code and method perform as expected, however we encourage the creation of higher fidelity simulations for further testing; consistency tests are also performed on the CFHTLenS data in Section \ref{Systematic Tests}. 

\subsection{Parameter Estimation}
The cosmological parameter estimation that we will present is performed using a Monte-Carlo Markov Chain (MCMC) algorithm 
that uses a Gaussian proposal distribution calculated using the Fisher matrix for the CFHTLenS survey, which uses the same 
posterior redshift information, survey masks, and ellipticity distributions calculated using a fiducial 
cosmology centred on WMAP7 (Komatsu et al., 2011) maximum likelihood values. This proposal distribution is efficient because 
parameter degeneracies are correctly included, however it does assume Gaussianity so no curvature in the likelihood surfaces is 
captured. The Fisher matrix used is given in Kitching, Heavens, Miller (2011). We create MCMC chains of $\gs 10^4$ evaluations, and 
create two chains per cosmology and CFHTLenS field in order to evaluate the Gelman \& Rubin (1992) statistic
(see Verde, 2007 for a clear explanation of this test), which we find to be consistent with 
the chains having converged for all results presented here\footnote{MCMC chains are available on request.}.

\section{Data}
\label{Data}
The CFHTLenS data and catalogue products are described in Erben et al. (2013), Heymans et al. (2012), Miller et al. (2013) and 
Hildebrandt et al. (2012). In this analysis we use all four wide fields (W1, W2, W3, W4) and 
reject pointings based on the Heymans et al. (2012) systematic tests: the tests were  
performed using a 2D correlation function method which should result in a more stringent rejection 
because the tests will be sensitive to contaminating effects from all scales, whereas in 3D cosmic shear we 
reject the smallest scales explicitly from the analysis. We use the 
catalogues presented in Erben et al., (2013) (the CFHTLenS catalogue) with the 
shape measurement ellipticities and weights described in Miller et al. (2013), created 
using \emph{lens}fit (Miller et al., 2007; Kitching et al., 2008), and the 
posterior redshift information described in Hildebrandt et al. (2012), created using {\sc BPZ} (Ben{\'{\i}}tez, 2000), 
which were tested for model-fidelity in Benjamin et al. (2012). 

The Benjamin et al. (2012) results show that the posterior $p_g(z)$ are 
consistent with the redshift distributions of galaxies with known spectroscopic redshifts, and with 
redshift distributions reconstructed from a cross-correlation analysis of $6$ photometric redshift bins; we 
also refer the reader to Figure 5 of Hildebrandt et al. (2012). From this analysis 
we infer that the set of galaxy template models used in Hildebrandt et al. (2012) are sufficiently complete such that the $p_g(z)$ 
are accurate representations of the true photometric error distribution in the range $0.2\leq z_{\sc BPZ} \leq 1.3$.
Furthermore Benjamin et al. (2012) based their analysis upon correlation functions. 
The higher sensitivity of 3D cosmic shear to redshift-dependent effects may be taken to imply that the method 
would be more sensitive to 
biases in photometric redshifts, however Kitching et al. (2008) find a similar required error on a global 
bias for 3D cosmic shear to requirements for weak lensing power spectrum tomography (e.g. Ma et al., 2005).
Although the $p_g(z)$ were tested in detail in Benjamin et al. (2012) and Hildebrandt et al. (2012), the selection
of late-type galaxies to avoid intrinsic alignment in this paper (see Section 3.1) 
may impact the applicability of those results. However, given the relatively large errors bars, we do 
not expect this to be significant for this study. 

\subsection{Galaxy Selection}
We make a redshift selection of posterior redshift distributions $p_g(z)$ of those 
galaxies with maximum-posterior values 
between $0.2\leq z_{\sc BPZ}\leq 1.3$. This is based on the cross-correlation analysis in 
Benjamin et al. (2012) who found consistency between spectroscopic and narrow band number 
densities and the summed $p_g(z)$ over these ranges, which is taken as evidence of a low-level 
of infidelity due to model/template-error in {\sc BPZ}. These galaxies have a weighted median 
redshift of $z_m=0.7$, and a mean effective number density of 11 galaxies per square arcminute 
over all fields before any selection. 

We make a cut in galaxy type by excluding all galaxies classified as
early-type, with a {\sc BPZ} type-parameter $T_B\leq 2$. The aim of this cut is to create a
model-independent removal of galaxies with a large intrinsic alignment signal, based on
the analysis of Heymans et al. (2013). However 
since the linear alignment model (Hirata \& Seljak, 2008) was used in that analysis there is some model-dependence. 
Mandelbaum et al. (2011) found similar results to Heymans et al. (2013). We will investigate a more sophisticated
3D intrinsic alignment removal in future work (see Merkel \& Schaefer 2013, for a theoretical study of intrinsic alignments within 
the 3D cosmic shear context). 

We use the image masks provided by Erben et al. (2013), and exclude those galaxies with 
$MASK\geq 2$ as described in Erben et al. (2013). We make no other selection of galaxies in 
the CFHTLenS catalogue. After selection the mean effective number density is 
$7$--$8$ late-type galaxies per square arcminute over all fields. 

\subsection{Scales}
\label{Scales}
We test two different ranges of radial scale  
$k_{\rm min}=0.001 h$Mpc$^{-1}$ and $k_{\rm max}=1.5 h$Mpc$^{-1}$ or $5.0h$Mpc$^{-1}$. 
The maximum radial scale of $1.5h$Mpc$^{-1}$ is defined to avoid the highly non-linear regime where baryonic effects are expected to become important
(see for example White, 2004; Zhan \& Knox, 2004; Jing et al., 2006; Zenter et al., 2008; Kitching \& Taylor, 2011; van Daalen et al., 2011; 
Yang et al., 2013; Semboloni et al. 2011, 2013). 
The {\sc halofit} (Smith et al., 2003) predictions also become unreliable at a redshift-dependent $k_{\rm max}$ 
(Giocoli et al., 2010) at similar scales. The higher scale of $5.0h$Mpc$^{-1}$ will enable a testing of these assumptions, probing the regime 
where feedback may be important. 

\begin{figure}
\psfig{file=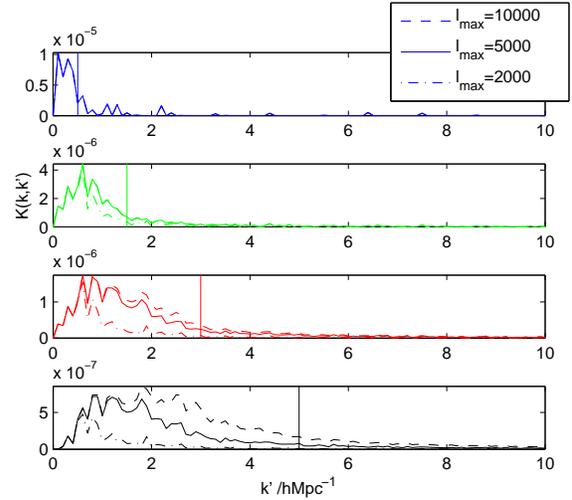, width=\columnwidth}
\caption{The 3D cosmic shear power spectrum kernel. The four panels show the kernel with which the matter power spectrum is convolved for several $k$-mode values 
(shown by vertical lines at values $0.5$, $1.5$, $3.0$ and $5.0 h$Mpc$^{-1}$) in the 3D cosmic shear power. For example the range of $k$-modes sampled in the matter power spectrum 
by a $k=0.5h$Mpc$^{-1}$ value in the 3D cosmic shear power spectrum are shown in the top panel. The range of $k$-modes sampled depends on the maximum $\ell$-mode used, and we show the kernel for 
three values $\ell_{\rm max}=2000$, $5000$ and $10$,$000$. In the cosmological analysis we use $k_{\rm max}=1.5h$ or $5.0h$Mpc$^{-1}$ and $\ell_{\rm max}=5000$.}
\label{kernel}
\end{figure}

The 3D cosmic shear power spectra probe particular scales in the matter power spectra; the maximum $k$-modes that we refer to are 
cuts in the data vector but do not probe only those physical scales. To quantify this we can re-write the $k$-diagonal part of the lensing power spectrum 
in terms of a kernel $K$ that acts on the matter power spectrum
\be 
C^{S}_{\ell}(k,k)=\int P(k';z) K(k',k){\rm d}k'
\ee
where $K(k',k)$ can be inferred in a comparison with equation (\ref{cov}). In Figure \ref{kernel} we show this kernel for several different values of $k$ (for the 
$W1$ field and using a WMAP7 cosmology; see Sections \ref{Data} and \ref{Results}) for several 
different maximum $\ell$ values. It can be seen that the kernel is, to a good approximation, confined to the region $k'\ls k$. The peak of the kernel is 
at lower $k'$ values as is expected because of the lensing effect being an integration along the line of sight. 
At $k\sim 5h$Mpc$^{-1}$ the impact of a maximum $\ell$-mode can be seen; that results because higher $\ell$-modes sample higher $k$-modes in general.  
In the analysis we also use $k'\gs 10h$Mpc$^{-1}$ in the integrations. 

This analysis is much more conservative than the 
correlation function analyses of the same data (Kilbinger et al., 2013; Heymans et al., 2013; 
Simpson et al., 2013; Benjamin et al., 2013) where for example
a minimum scale of $0.8$ arcmin is used, which is equivalent to a data vector cut of $k\leq 27$,$000/r(z)\ls 30h$Mpc$^{-1}$ 
for the closest redshifts, or $k \ls 10h$Mpc$^{-1}$ at the median redshift\footnote{See Benjamin et al. (2013) Section 4 for a more thorough 
discussion of this $k$-cut for correlation function analyses; however much larger $k$-modes will contribute to the interpretation of these 
because of the very broad kernel used in this analysis (Bessel functions $J_{0,4}[\ell\theta]$).
Here we quote the maximum scale in the data vector that is used in the likelihood function in such an analysis, 
for comparison with the $k$-cuts used in this paper.}. 
A more conservative correlation function analysis in Kilbinger et al., (2013) 
uses a minimum scale of $17$ arcmin in the data vector but the remaining modes still necessarily contain a mix of information from all physical scales; this is a because for a correlation function method the kernel, 
a Bessel function $J_{0,4}(\ell\theta)$ has significant power at all scales, 
so a cut in the data vector at a particular scale does not translate into a cut in a physical scale. 

For the angular scales we use $\ell_{\rm min}=360$ to avoid any residual systematic effects on scales larger than a single CFHTLenS pointing. 
To sample the 2D $\ell$-space we then create modes $\ell_x=i\ell_{\rm min}$ and $\ell_y=j\ell_{\rm min}$ 
(where $i$, $j\in{\mathbb Z}$) such that the 
magnitude of the $\ell$ vector $\ell=\sqrt{\ell_x^2+\ell_y^2}$ is always less than 
$\ell_{\rm max}=5000$. We use integer multiples of $\ell_{\rm min}$ for computational reasons, but note that this could limit the signal-to-noise of the final 
results. We evaluate the data vectors and the theoretical covariance at each point in this 
2D space, concatenating those combinations of $\ell_x$ and $\ell_y$ that give the same $\ell$, resulting in $164$ 
independent angular modes. We then sum the likelihood values, equation (\ref{el}), for parameter estimation. 
In the radial direction for every $\ell$-mode, we use $50$ $k$-modes linearly sampled between $0.001$--$5.0 h$Mpc$^{-1}$ (therefore
$15$ for the $k\leq 1.5h$Mpc$^{-1}$ cut).
The MCMC chain is common for all the data, where at each point the log-likelihood is summed over all fields. 

\subsection{3D cosmic shear power spectra}
\begin{figure*}
\psfig{file=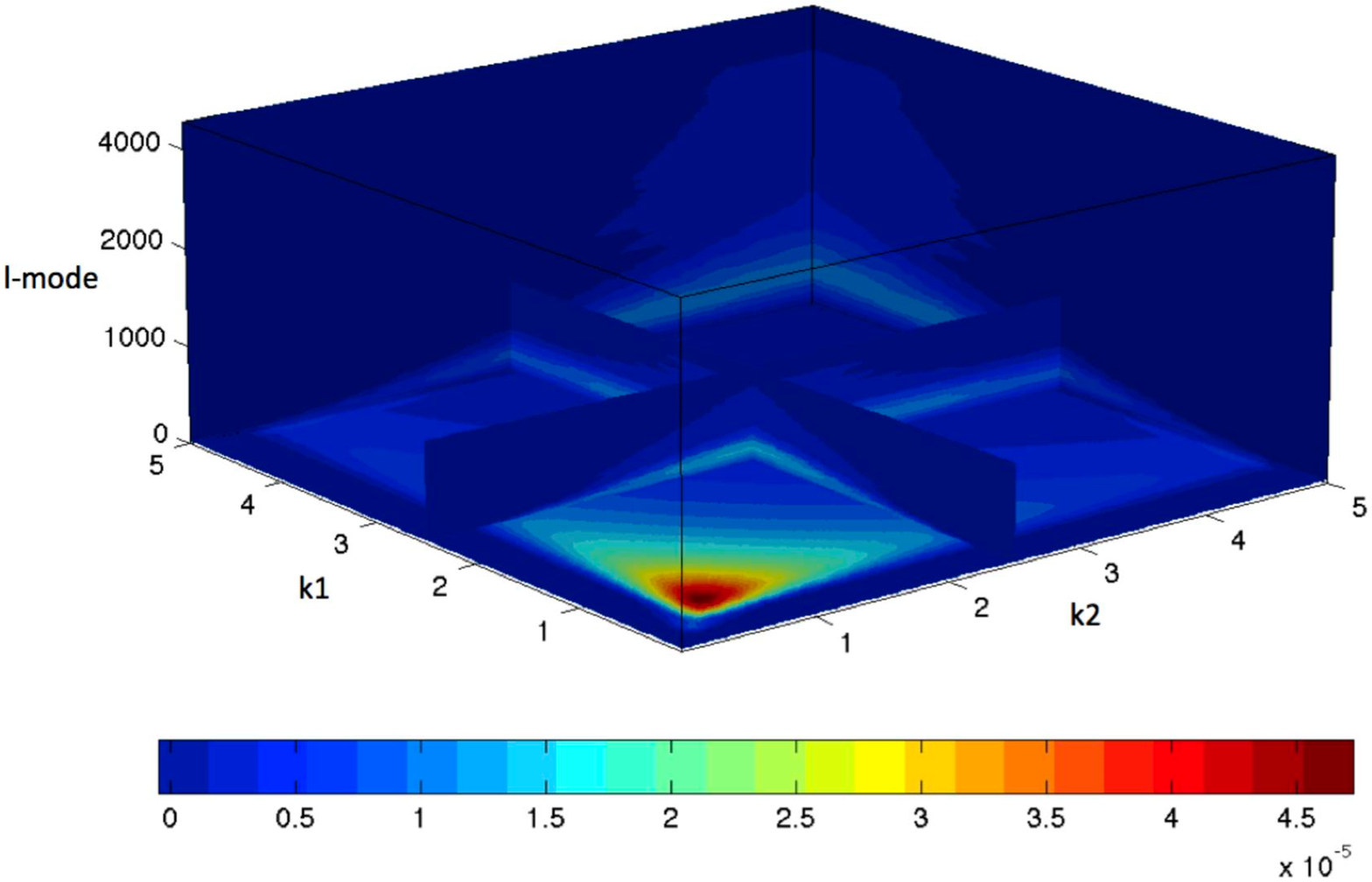, width=1.5\columnwidth, angle=-90}
\vspace{-0.5cm}
\psfig{file=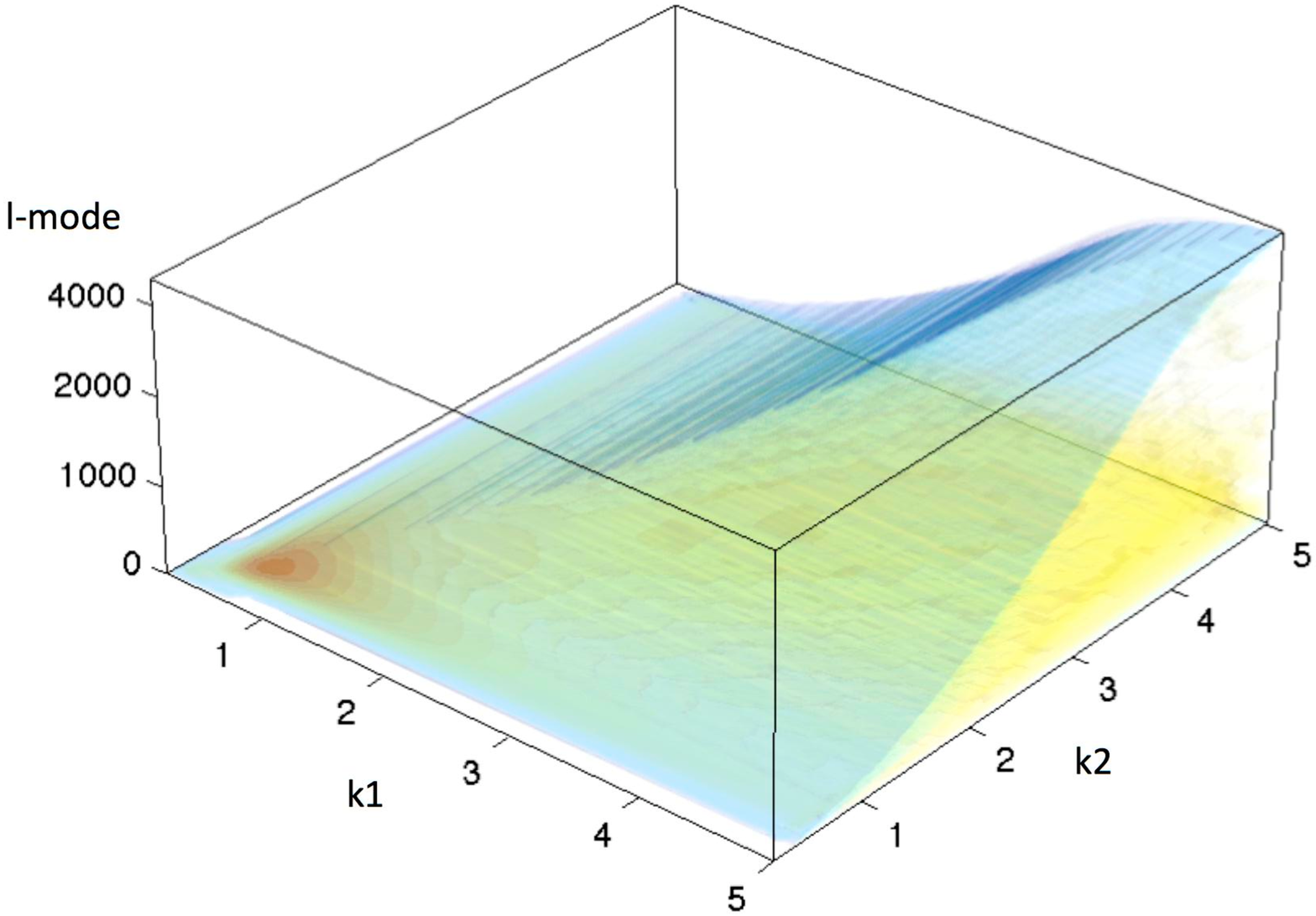, width=1.25\columnwidth, angle=90,clip=}
\caption{3D representations of the signal part of the 3D cosmic shear power spectra $C^S_{\ell}(k_1,k_2)$ (equation \ref{cov}) averaged over the real and imaginary parts. 
The upper panel shows a slice plot through the 3D ($\ell$, $k_1$, $k_2$) space 
plotted on the $z$, $y$ and $x$ axes respectively. The slices/cross-sections through the 3D cube in the top panel are at $k=2.5 h$Mpc$^{-1}$. The lower panel shows the same power spectrum but in an iso-surface represenation. The colour bar 
gives the amplitude of the power at each point in the 3D space for both panels.}
\label{3dplot}
\end{figure*}

The 3D cosmic shear power spectra are inherently complex three-dimensional objects in ($\ell$, $k_1$, $k_2$) space. In Figure \ref{3dplot} 
we show the signal, cosmology-dependent, part of the 3D cosmic shear power spectrum in this space. This shows the broad features that 
lower $\ell$-modes contain more power and that as the $\ell$-mode increases fewer $k$-modes are accessible because of the Bessel 
function behaviour, that we discuss further below.  

\begin{figure*}
\psfig{file=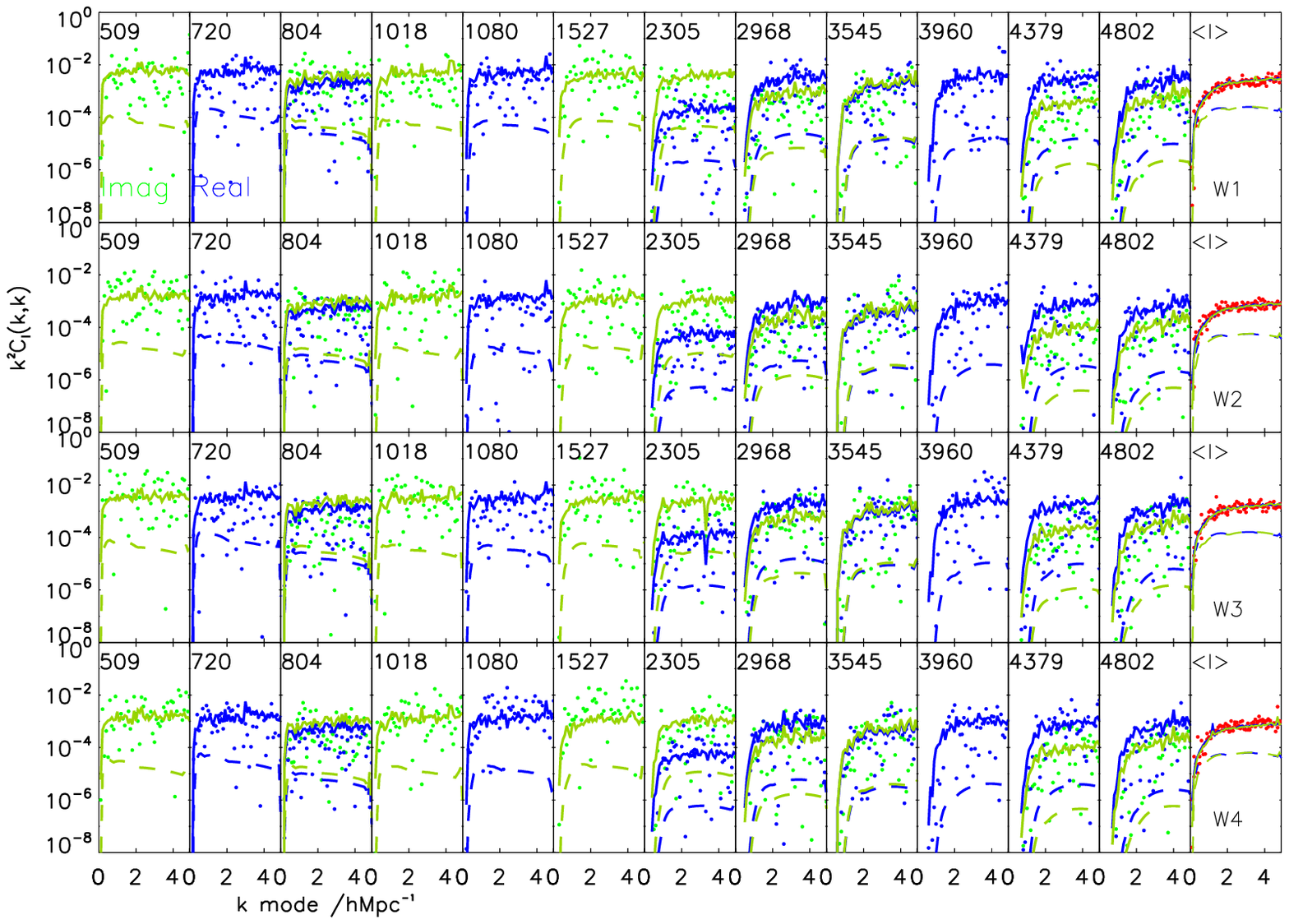, width=1.7\columnwidth, angle=0}
\psfig{file=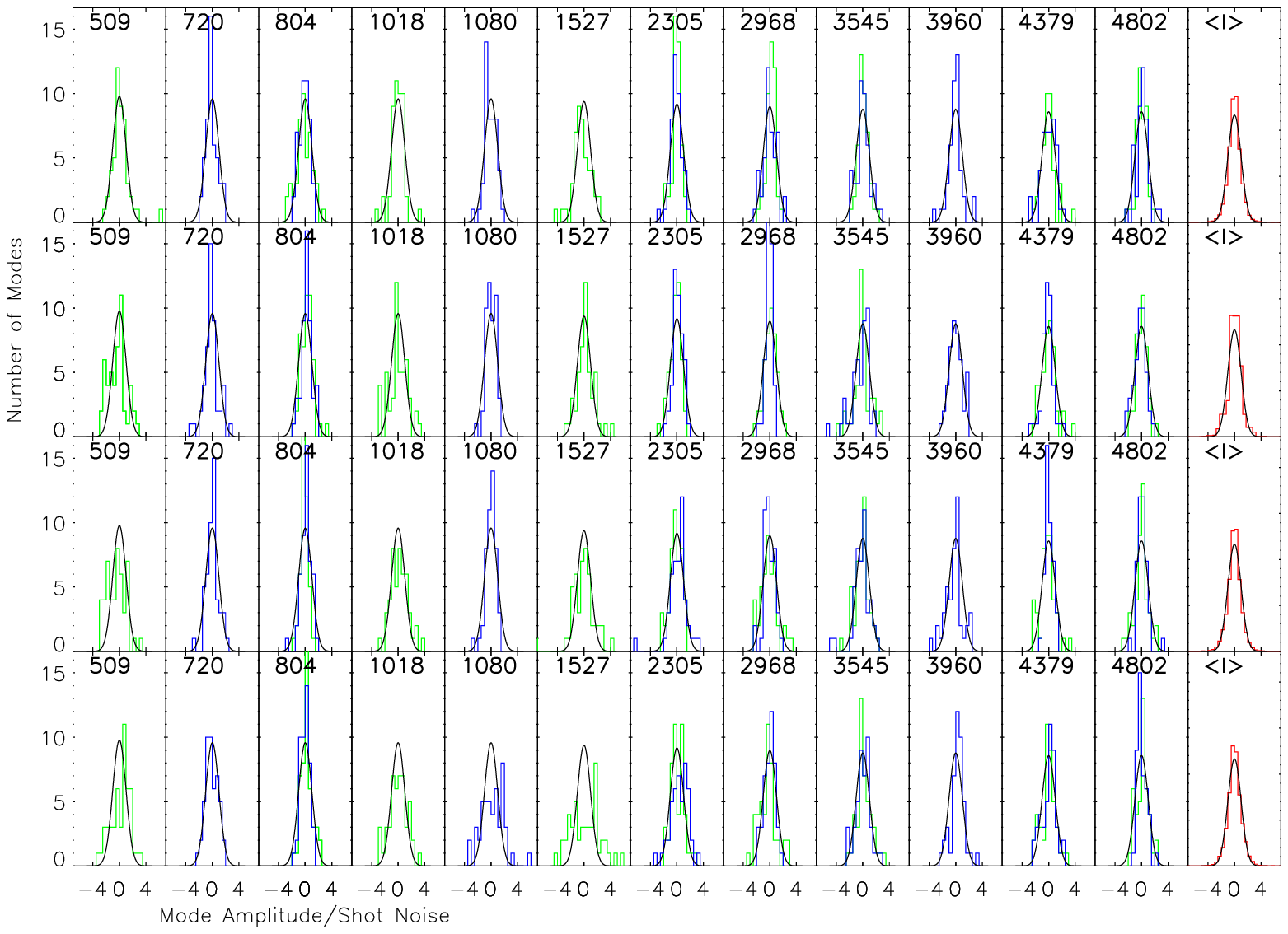, width=1.7\columnwidth, angle=0}
\caption{Upper panel: A $k$-diagonal cross-section through the 3D power spectra. The data points show the real and imaginary 
values of the transform 
coefficients squared $\gamma^2_E(k,\ell)$ as a function of $k$ for a representative set of $12$ $\ell$-modes from the $\sim 164$ $\ell$-modes computed for each 
CFHTLenS field: W1, W2, W3 and W4 respectively from top to bottom in rows. The green points and lines are for the imaginary 
part, and the blue for the real part. The dashed and solid lines
show the diagonal part of the signal and noise covariances respectively in the $k$ 
direction $C_{\ell}(k,k)$ calculated at a reference WMAP7 cosmology (Komatsu et al., 2011), for the imaginary (green) and 
real (blue) parts of the covariance. The 
rightmost column shows the mean of the same cross-section averaged over all $\ell$-modes for each field, and averaged over the real 
and imaginary parts. Lower panel: A histogram of the real (blue) and imaginary (green) shear coefficients for each $\ell$-mode in the upper panel, and also averaged over all $\ell$-modes, divided by the expected shot noise. The black lines show unit Gaussian distributions.}
\label{3dcs}
\end{figure*}

To present the full 3D power spectrum in a more accessible form we can take conditional cross-sections of this space or make projections. 
In Figure \ref{3dcs} we show, for a representative set of $\ell$-modes for each field, the diagonal 
part of the power spectra $C_{\ell}(k,k)$ and compare this with the square of $E$-mode data vector 
values for both the real and imaginary parts of the power spectrum; these quantities should be approximately equivalent if the power spectrum is nearly 
diagonal in the $k$ direction, we scale the quantities by $k^2$ so that one would have a flat spectrum if there 
were equal power in each shell in $k$ space. Not all $\ell$-modes have both real and imaginary power spectra because of the nature of the 
complex derivatives ($D_1$ and $D_2$ in equation \ref{cov}) in the $\ell$ coordinate system. 

The dominant feature that one sees in Figure \ref{3dcs} is the sharp drop in power at low $k$ for each $\ell$, 
which is expected and is due to the Bessel function behaviour $j_{\ell}(kr)\approx 0$ for $\ell\geq kr$. In this case for a given $k$ 
we expect to find power at $k\geq \ell/r_{\rm max}\gs \ell/(3000z_{\rm max}h^{-1}$ Mpc$)$. A further clear 
feature is that for any given ($\ell$, $k$) mode the signal-to-noise ratio of the power spectrum is 
much less than unity, typically $\sim 10$ per cent (the ratio of the dashed lines to the solid lines
in Figure \ref{3dcs}). We note, however, that in total over CFHTLenS we have 
$\sim 200\times 50\times 4\sim 40$,$000$ independent modes. This is expected for a survey of this 
size, but since the signal-to-noise increases linearly with the number of galaxies for a 
particular mode future surveys may even detect individual modes at signal-to-noise greater than unity. 
We also show the same cross-section in $k$ averaged over all $\ell$-modes. 
The consistency of the shot-noise part with the B-mode is in agreement with a similar conclusion reached 
in a mass-mapping analysis of the same data in van Waerbeke et al. (2012).    

\subsection{2D \& tomographic cosmic shear power spectra}
A further projection that one can make of the 3D power spectrum is to average over the $k$ direction 
to create a purely angle-dependent representation of the power. In Appendix D we show how one can compute 
such 2D, or tomographic, power spectra from the full 3D case. Using this formalism one 
could calculate any 2D auto-correlation or cross-correlation power spectrum between any pair of redshift 
bins. 

\begin{figure*}
\psfig{file=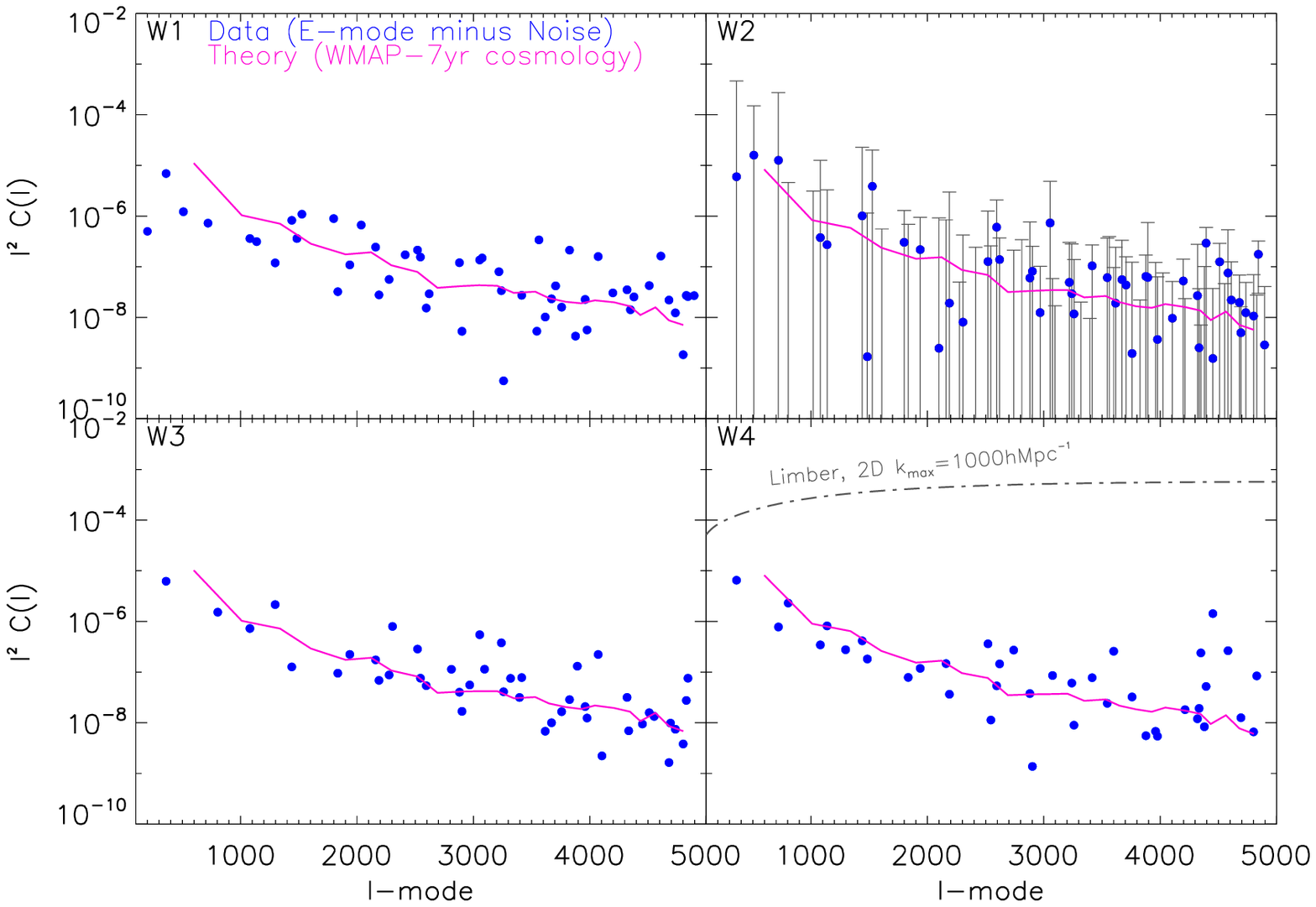, width=2\columnwidth, angle=0}
\caption{The projected 2D cosmic shear power spectrum for each of the CFHTLenS fields; computed by 
integrating the full 3D cosmic shear power. The data points show 
the $E$-mode only power as a function of $\ell$. The solid line 
shows the 2D power spectra estimates calculated using a reference 
cosmology of WMAP7 (Komatsu et al., 2011). For the W2 field we show the error bar on each point, which are also typical of the other fields. 
Because of the logarithmic $y$ axes negative values as a result of noise are not shown. 
For illustration, in the W4 field only, the grey dot-dashed line shown is 
the 2D cosmic shear power spectrum that one would compute from data evaluated on a plane, or 
from theory if no cut in the radial $k$ direction were imposed in the Limber-approximated calculation.}
\label{2dcs}
\end{figure*}

In Figure \ref{2dcs} we show the 2D projected power when averaging over the whole redshift range, a `2D 
cosmic shear power spectrum', for each of the CFHTLenS fields. We show the sum of $E$-mode power averaged over real and
imaginary parts with the shot noise subtracted and compare this to the 2D projected signal 
calculated with a reference WMAP7 cosmology (Komatsu et al., 2011). Other 2D power spectrum analyses for weak lensing have been presented by 
Pen et al. (2002) using the VIRMOS-DESCART survey, Brown et al. (2003) using the COMBO-17 survey, and Heymans et al. (2005) using the HST GEMS survey.
We do not combine the fields in our visualisations of the data because such a combination is not a necessity for our analysis; 
such a combination is also not trivial because the mixing matrices and number density vary between fields. The theory 
curves plotted are convolved with 
the mixing matrix, computed from the inhomogenous and non-smooth masks in the data, 
and hence are not smooth as may be expected if this was not done.

\begin{figure*}
\psfig{file=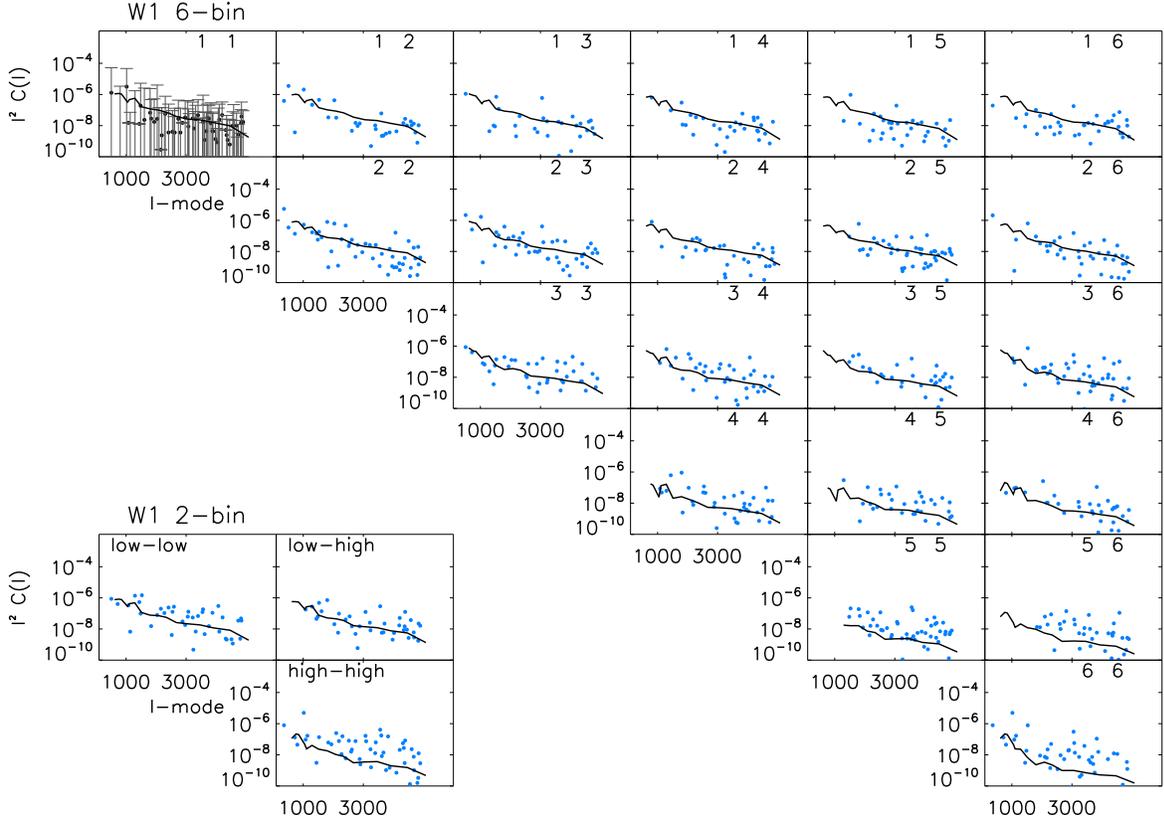, width=2\columnwidth, angle=0}
\caption{Tomographic power spectra for the CFHTLenS W1 field; computed by
integrating the full 3D cosmic shear power. The data points show
the summed $E$-mode power as a function of $\ell$. 
The solid lines show the power spectrum for a WMAP7 (Komatsu et al., 2011) 
cosmology. The main 21 panels show a 6-bin tomographic set, the numbers refer to the redshift bin combinations where bins 1 to 6 have redshift ranges 
$0.2<z\leq 0.39$, $0.39<z\leq 0.58$, $0.58<z\leq 0.72$, $0.72<z\leq 0.86$, $0.86<z\leq 1.02$, $1.02<z\leq 1.30$ respectively (the same 
as Heymans et al., 2013 who present correlation function tomography); the diagonal panels show the intra-bin power spectra, 
the off-diagonals the inter-bin power spectra. The smaller set of 3 panels show a 2-bin tomographic set, the labels low and high refer to the redshift 
bin combinations where low and high redshift ranges are $0.5<z\leq 0.85$ and $0.85<z\leq 1.30$ respectively (the same as Benjamin et al., 2012 who present 
correlation function tomography). In the first panel we show the error bars on each point, which are typical for the other bins both sets.
Because of the logarithmic $y$ axes negative values as a result of noise are not shown.} 
\label{6bcs}
\end{figure*}

In Figure \ref{6bcs} we show a set of tomographic power spectra using both $2$ and 
$6$ redshift bins, using the same redshift binning used in Benjamin et al. (2012; 2-bin) and Heymans et al. (2013; 6-bin) for correlation function analyses. 
We show the intra-bin or `auto-correlation' power spectrum for each redshift bin and the inter-bin or `cross-correlation' power spectrum for each redshift bin 
combination in the set. It can be seen that the scaling of the data with redshift matches that expected from a WMAP7 cosmology (Komatsu et al., 2011); 
except in the highest redshift bin where we see some excess of power. There is more 
power at high-$\ell$ for high redshifts, which is what one expects, the drop in power is seen at
$\ell_{\rm max}\approx k_{\rm max}r(z)$. The overall drop in power from low to high $\ell$ is due to the maximum $k$ cut similar to the 
2D case (Figure \ref{2dcs}). The smaller amount of inter-bin power, decreasing as the bin separation increases, is also expected as 
there is less common lensing material between the bins.  
Note that this presentation could be extended to an arbitrary number of 
bins\footnote{However ultimately limited by the number of galaxies in the survey. Decreased photometric redshift precision would increase correlation between 
bins but would not limit the number of bins.}. Because the signal-to-noise of the
shear transform coefficients is low in CFHTLenS (see Figure \ref{3dcs}) the projected power spectra also have a high level of noise; as a consequence 
points that are scattered to negative values as a result of noise are not shown in Figures \ref{2dcs} and \ref{6bcs}, and we show typical error bars for each point. 

If one does not use 3D cosmic shear but instead uses 2D or tomographic approximations then 
it is important to correctly account for the effect that 
any cuts made in the radial $k$-modes have on the angular $\ell$-modes used in both the theoretical calculation 
and in the measurement of the power spectra from the data. 
Cuts in the radial $k$-modes on the matter power spectrum for example 
have, as a result of the projections and the Limber approximation, an impact on $\ell$-modes in the regime 
$k\geq \ell/r_{\rm max}\gs \ell/(3000z_{\rm max}h^{-1}$ Mpc$)$. Cuts on radial $k$-modes therefore result 
in a suppression of 2D or tomographic power at a fixed $\ell$-mode as power from the cut modes is removed. 
This effect is readily computable from theory, either from the full 3D power spectrum 
or by making tomographic approximations (e.g. Hu, 1999; Kitching, Heavens, Miller, 2011), 
and indeed it is necessary to do so because small 
scales should be removed due to uncertainties in baryonic feedback, and the non-linear power spectrum.  

However, the estimation of a 2D power spectrum from data, which consistently removes these modes for a
correct comparison to theory, has not been 
demonstrated until now. 
In fact the computation of 2D power, from an inherently 3D field, 
on a plane will contain contributions from all $k$-modes. In Figure \ref{2dcs} we represent what 
one would have computed from data using this procedure, assuming the Limber approximation with no $k$-mode cut, 
with the grey dot-dashed line: in this case the data would be 
orders of magnitude away from the theoretical predictions. One can mitigate this by computing the 
theoretical power to larger $k$ values, but such a procedure carries uncertainties. 
If one projects the shear field on to a plane then both the selection function of the galaxies and a correct
removal of power from $\ell\geq k_{\rm max}r(z)$ would have to be performed. 
Alternatively one can use the full 3D transform coefficients, and use the projection presented in this paper to remove $k$-modes from the 
data covariance. A further point is that the $k$-mode cuts in the data vector translate to a particular kernel with which the matter power 
spectrum is convolved as discussed in Section \ref{Scales}.

\subsection{Systematic Tests}
\label{Systematic Tests}
\begin{figure*}
\psfig{file=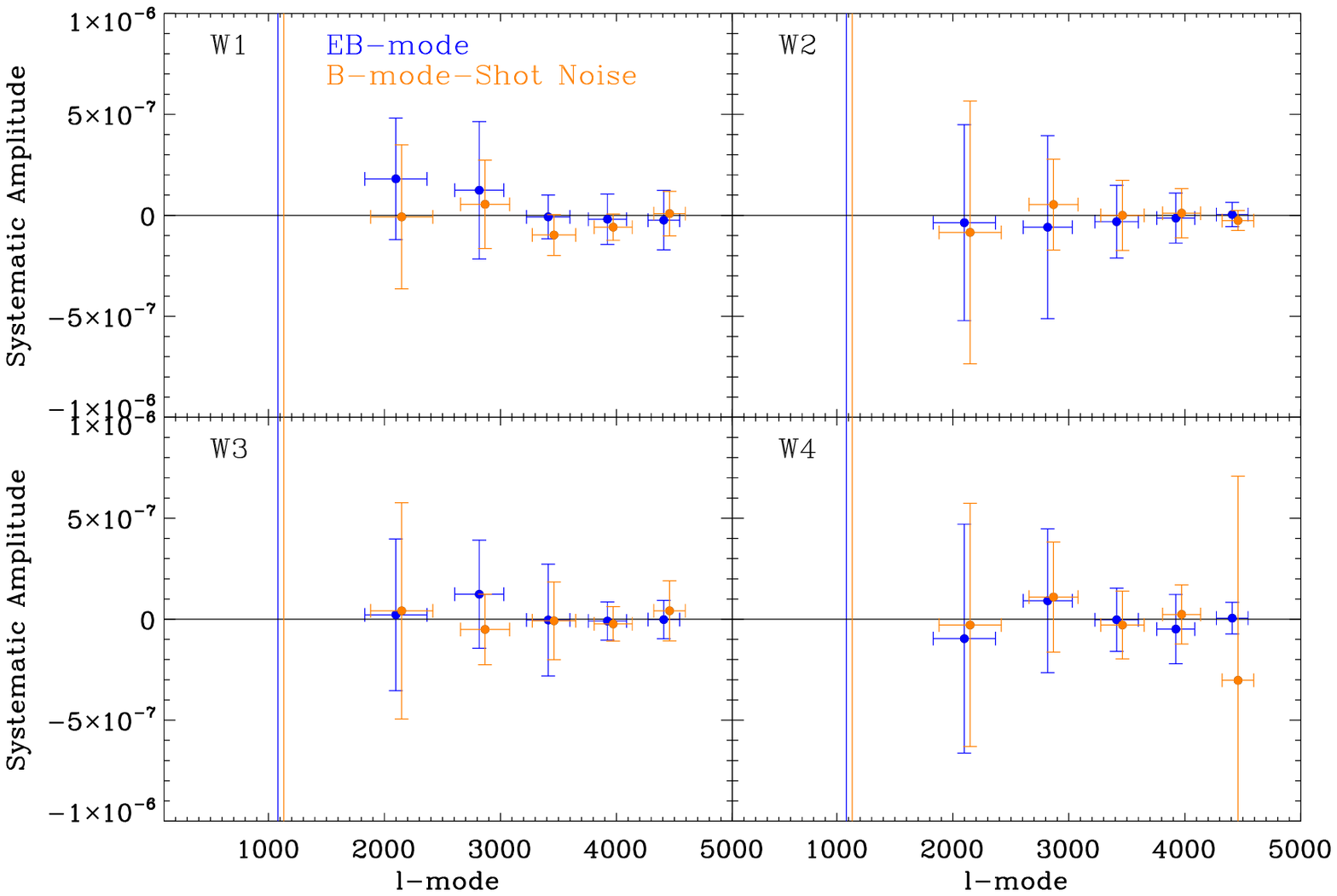, width=1.75\columnwidth, angle=0}
\psfig{file=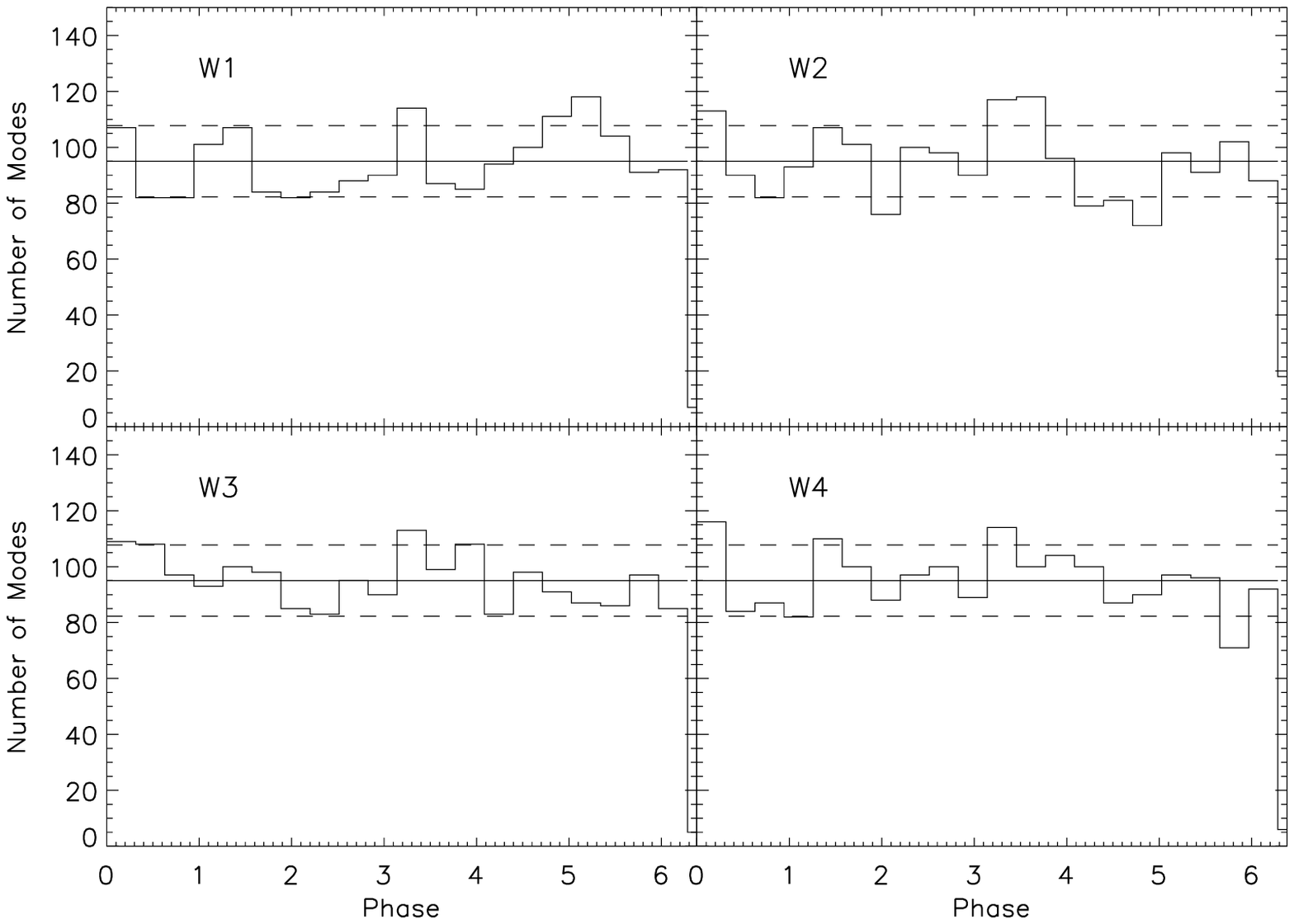, width=1.75\columnwidth, angle=0}
\caption{Upper panels: For each of the CFHTLenS fields we show the shot noise subtracted B-mode power (blue points) and the EB cross power spectra (orange points) 
Each of these should be consistent with zero. We have binned the $\ell$-modes into $6$ bins and show the error bar associated with each; we have shifted the $B$-mode 
by a small amount away from the bin center (used for the EB points) in $\ell$ for clarity in plotting. 
Lower panels: For each of the CFHTLenS fields we show a histogram of the complex phase of the observed shear coefficients (equation \ref{ephase})
Each of these should be consistent with a uniform distribution over the range $[0, 2\pi]$, if the data is isotropic. The solid horizonatal line shows the 
expected mean number of modes per bin, and the dashed lines show the expected $1\sigma$ error.}
\label{systest}
\end{figure*}

There are several systematic tests that one can perform, under particular assumptions, to determine whether the 
power spectra calculated from the data are consistent with expectations: 
\begin{itemize}
\item 
The B-mode part of the power spectrum should be consistent with shot noise only (equation \ref{noi}), because cosmic shear only 
induces E-mode power. Therefore the B-mode power minus the expected shot noise power spectrum should be consistent with zero. 
This assumption can break down due to intrinsic alignments (see e.g. Merkel \& Schaefer, 2013), but at the level of precision attainable from 
CFHTLenS, and the fact that we remove the galaxies that are most likely to be contaminated with intrinsic alignments, this is a valid systematic test.
\item 
The cross power spectrum between the E and the B-mode power should be consistent with zero. A non-zero E-B power spectrum would correspond to 
a mixing of E and B-mode power which is expected to be zero, except in some exotic cosmologies (see Amendola et al. 2013 for a review), or as a result of residual 
systematic B-mode power being mixed with the E-mode power through the application of the mixing matrix. 
\item 
For a Gaussian random field the phase of the E and B-mode power spectra for a given mode is  
\be
\label{ephase}
\phi={\rm atan}\left(\frac{{\mathbb I}[\gamma(k,\ell)]}{{\mathbb R}[\gamma(k,\ell)]}\right). 
\ee
The distribution of phases should be random, and consistent with a uniform distribution over $[0, 2\pi]$ (see Coles et al. 2004 for a study of phases in a CMB study) 
if there is no prefered direction in the data (this tests sensitivity to a shift in the origin of the coordinate system used). 
The shear coefficients used in the above equation are the observed shear coefficients (equation \ref{obscoeff} in Appendix A) to test the isotropy of the on-sky 
shear field.  
\end{itemize}
We show the result of the first two of these systematic tests in Figure \ref{systest} for each of the four fields as a function of $\ell$, averaged over $k$, and the real 
and imaginary parts of the power spectra. We find that as expected each of these tests is consistent with zero. 
In Figure \ref{systest} we also show the distribution of the complex phases of the observed transform coefficients averaged over all $\ell$ and $k$-modes, which 
we find to be consistent with a uniform distribution for each field. 

\section{Results}
\label{Results}
We now present the cosmological parameter constraints found from 3D cosmic shear applied to the CFHTLenS data. 
The cosmological parameter set we use is a waCDM set with $\Omega_{\rm M}$, $\Omega_{\rm B}$, $\sigma_8$, $h$, $w_0$, $w_a$, $n_s$  
with others fixed at WMAP7 maximum likelihood values (Komatsu et al., 2011), we also assume flatness 
i.e. $\Omega_{\rm DE}=1-\Omega_{\rm M}$, and a sum of neutrino mass of zero. 
We also consider a wCDM parameter set where $w_a=0$, and a LCDM parameter set where $w_0=-1$ and $w_a=0$.  
Constraints on other cosmological 
parameters are expected to be dominated by CMB constraints for this size of lensing survey, except 
possibly the neutrino mass. The dark energy 
equation of state is parameterized using a Taylor expansion in scale factor such that $w(z)=w_0+[z/(1+z)]w_a$. 

\subsection{Priors}
We will present the 3D cosmic shear parameter error in combinations with priors from previously analysed cosmological 
data sets. These are: 
\begin{enumerate}
\item 
\emph{Planck}: We include constraints from the \emph{Planck} 1st year data. See Planck (2013) and the PLAIO\footnote{\url{http://pla.esac.esa.int/pla/aio/planckProducts.html}} for a description of the data products. 
We use the \emph{lowl\_lowLike} chains and for the waCDM parameter set use the combination of Planck+BAO. 
\item 
$H_0$: The constraint on the dimensionless Hubble parameter $h=H_0/(100 {\rm kms^{-1}Mpc^{-1}})=0.738\pm 0.024$ 
from Riess et al. (2011). We apply this assuming a Gaussian prior distribution.  
\item 
WMAP7+SN+BAO: We include results from Komatsu et al. (2011) for the CMB in combination with priors used in that analysis\footnote{We use WMAP7 for consistency with other CFHTLenS results, and \emph{Planck} for current constraints, we also refer the reader to WMAP9 (Hinshaw et al., 2013) for a further CMB data set that could also be used.}. We use the MCMC chains made available subsequently\footnote{Available here 
\url{http://gyudon.as.utexas.edu/~komatsu/wmap7/} and here \url{http://gyudon.as.utexas.edu/~komatsu/wmap7/wacdm+lz/wmap7+h0+snconst/} for the waCDM parameter set 
and from here \url{http://lambda.gsfc.nasa.gov/product/map/dr4/parameters.cfm} for the wCDM parameter set.}
that also include information from the Hicken et al. (2009) supernovae data set (+SN) and BAO information from Percival et al. (2010) (+BAO). 
\end{enumerate}
The WMAP7+SN priors we use do not contain systematic errors on the supernovae constraints. This is addressed in  Conley et al. (2011) 
who find that the combined 
WMAP7+SN constraints including systematic are not biased with regard to Komatsu et al. (2011) 
(due to the orthogonality of the CMB and SNIa contours even when including systematics), but that uncertainties from 
SN alone are increased by a factor of $2$. Komatsu et al. (2011) included an $H_0$ prior from Riess et al. (2009) of $h=0.742\pm0.036$, and we modify the weights of 
the WMAP7 MCMC chains to remove the Riess et al. (2009) $H_0$ prior and  
include the Riess et al. (2011) $H_0$ prior for this paper.
In addition we include some physical priors $\Omega_{\rm M}>0$, $\Omega_{\rm B}>0$, $h>0$, $\sigma_8>0$ to prevent the MCMC chains 
from moving into unphysical parts of parameter space. 
We also include i) the same uniform priors as Kilbinger et al. (2013) of $\Omega_{\rm B} \in [0.0; 0.1]$, $n_s \in [0.7; 1.2]$, 
and ii) some priors that result from the stability of {\sc camb}, where we exclude the 
the ranges $\Omega_{\rm M}<0.05$, $h<0.1$ and $(w_0>-0.5 \wedge w_a>0.8)$ (see Appendix E). 

\subsection{Scales}
\label{Scales}
\begin{figure*}
\psfig{file=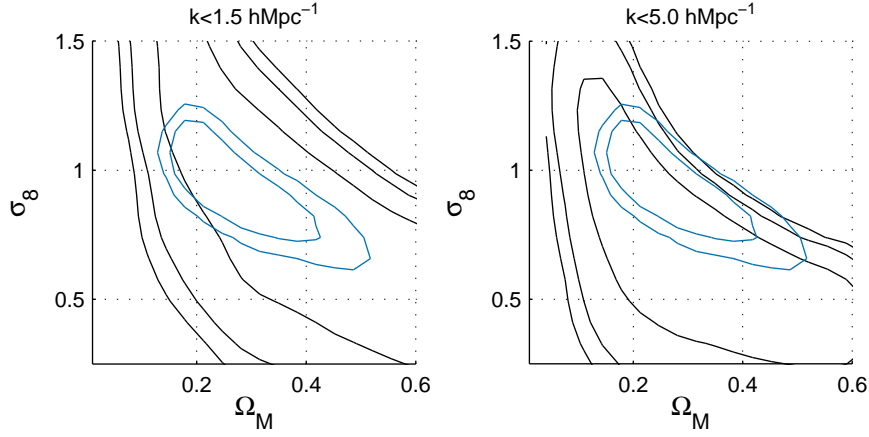, width=1.4\columnwidth, angle=0}
\caption{Constraints in the $(\sigma_8,\Omega_{\rm M})$ plane for a wCDM cosmology as a function of the range in $k$-modes used in the analysis (black contours). 
We show the wCDM WMAP 7 yr contours (blue inner lines) for comparison. Contours shown are 2-parameter $1$ and $2\sigma$ confidence regions. In the lefthand 
panel the $k$ range is $k\leq 1.5h$Mpc$^{-1}$; and in the righthand panel for $k\leq 5.0h$Mpc$^{-1}$}
\label{scales}
\end{figure*}

In Figure \ref{scales} we show constraints in the $(\sigma_8,\Omega_{\rm M})$ plane for two different ranges in scale. 
In this projection we find that results are consistent when the maximum $k$ is increased from $1.5$ to $5.0h$Mpc$^{-1}$, but 
we will see tension later when combined with \emph{Planck}. Moreover even over 
this small change in scale a lower $\sigma_8$ is preferred as the maximum $k$ increases. We show numerical results for a 
fit to the function $\sigma_8(\Omega_{\rm M}/0.27)^{\alpha}$=constant in Table \ref{comparison} where we find that the normalisation 
is affected by a change in scale but that the slope is unaffected.
\begin{figure*}
\psfig{file=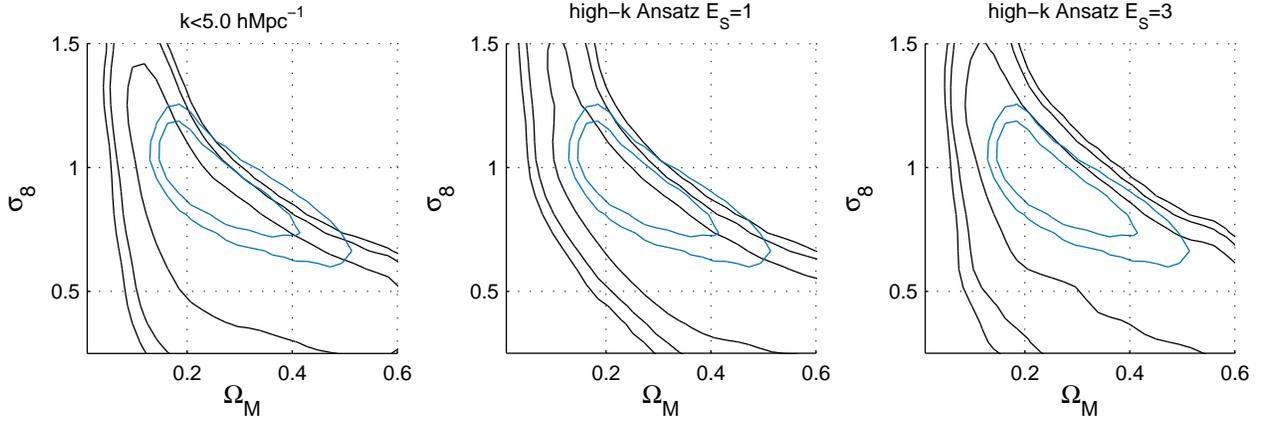, width=2.0\columnwidth, angle=0}
\caption{Constraints in the $(\sigma_8,\Omega_{\rm M})$ plane for a wCDM cosmology for $k\leq 5.0h$Mpc$^{-1}$, 
including the functional ansatz from van Daalen et al. (2011) and 
Semboloni et al. (2011, 2013) for the 
effect of baryonic feedback on the matter power spectrum, that we parameterise in equation (\ref{bs}). The lefthand panel reproduces the plot from Figure \ref{scales} 
for comparison; the middle panel shows the constraints using the functional ansatz predicted, with no amplitude change $E_S=1$; 
the righthand panel shows the constraints when using a power spectrum that is damped three times more than that predicted $E_S=3$. We show the
wCDM WMAP 7 yr contours (blue inner contours) for comparison.}
\label{bars}
\end{figure*}

One possible explanation for the preference of lower power above $k=1.5h$Mpc$^{-1}$ is that the impact of baryonic feedback on the total 
matter power spectrum is being seen. van Daalen et al. (2011) and Semboloni et al. (2011, 2013) used 
N-body simulations to investigate the impact of baryons (via AGN feedback) 
on the matter power spectrum and provided a functional ansatz for their predicted effect: we refer to the solid blue lines in Semboloni et al. (2011, Figure 1) and 
Semboloni et al. (2013, Figure 5), such that we parameterise the total matter power spectrum as 
\be 
\label{bs}
P(k,z)=\left[E_S\left(\frac{P_{\rm B}(k)}{P_{\rm N}(k)}-1\right)+1\right]P_{\rm DM}(k,z)
\ee
where $P_{\rm DM}(k,z)$ is the original ({\sc camb}) power spectrum as a function of scale and redshift; $P_{\rm B}(k)/P_{\rm N}(k)$ is the functional 
form for the ratio of the total matter to dark matter power spectra from Semboloni et al. (2011), which we assume to be redshift-independent over 
scales $k\leq 5h$Mpc$^{-1}$ and the redshift range of CFHTLenS; 
$E_S$ is an additional parameter that controls the amplitude of the damping term. 
As a partial test of the impact of baryons on the cosmic shear power spectra we show in Figure \ref{bars} the $(\sigma_8$, $\Omega_{\rm M})$ plane using 
matter power spectra of the form given in equation (\ref{bs}) for $E_S=1$ and $3$; these two values are meant to be representative of the plausible range of 
suppression and encapsulate the original suggested form and a suppression below which there could be zero power at some scale, it should be noted however that 
the upper value of $3$ is much more extreme than that expected which is likely to be in the range $0$--$1$. 
We find as expected that the preferred value of $\sigma_8$ increases as the modelled 
matter power spectrum is damped. This is not direct evidence of the impact of baryons on the power spectrum, but we do find better consistency between the 
$k\leq 1.5h$Mpc$^{-1}$ and $k\leq 5h$Mpc$^{-1}$ ranges when the baryon functional ansatz is included in the modelling. This 
is a functional ansatz only, where for example 
redshift dependence is not included; however redshift independence is not a very bad approximation since AGN activity peaks at early times\footnote{However 
it is not clear whether material blown out by AGN activity may, or may not, be able fall back into its original environment.}. 
The feedback behaviour should be noted in the interpretation of the results of 
all cosmic shear cosmological constraints that use scales $k\gs 1.5h$Mpc$^{-1}$. 

\subsection{wCDM cosmologies} 
Here we explore the combination of CMB constraints with those from lensing in comparison with similar combinations 
from other cosmological 
probes in the wCDM parameter space. The CMB alone (not accounting for lensing of the CMB) suffers 
from a geometric degeneracy which means the constraints are large in particular parameter directions, in particular for
$\Omega_{\rm M}$, $h$ and $w$. CMB measurements alone can lift the degeneracy to some degree with CMB lensing and the ISW effect, 
but are generally combined with other cosmological probes in order to take full advantage of the statistical power of additional datasets.
 
\begin{figure*}
{\bf $k_{\rm max}\leq 1.5h$Mpc$^{-1}$}
\psfig{file=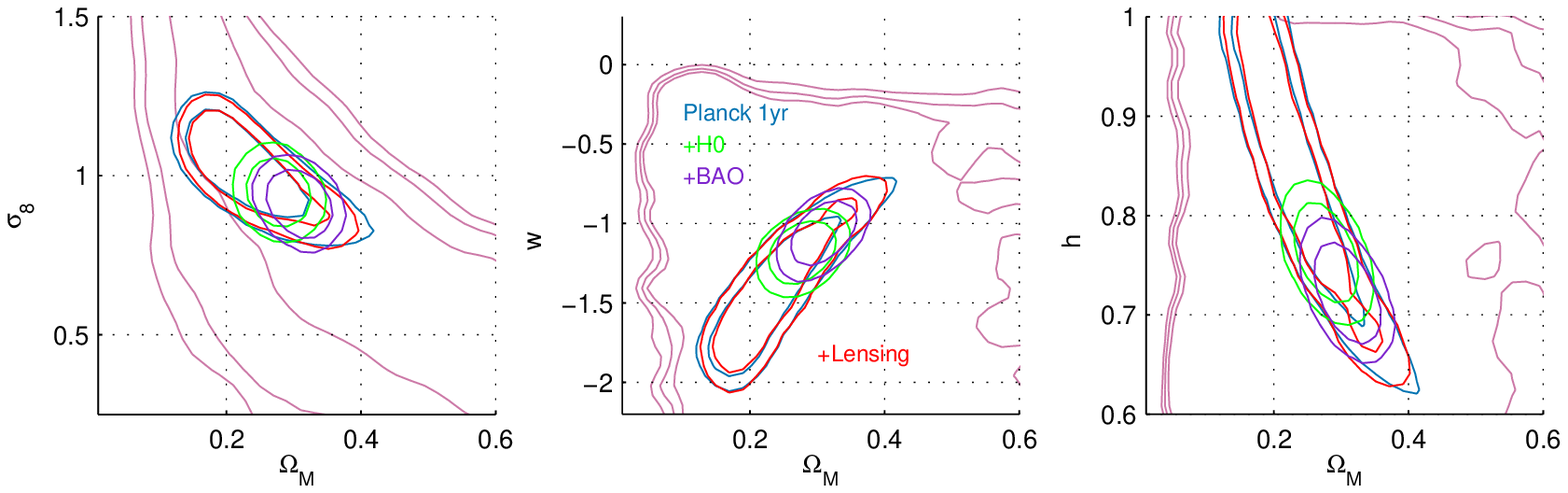, width=1.8\columnwidth, angle=0}
{\bf $k_{\rm max}\leq 5.0h$Mpc$^{-1}$}
\psfig{file=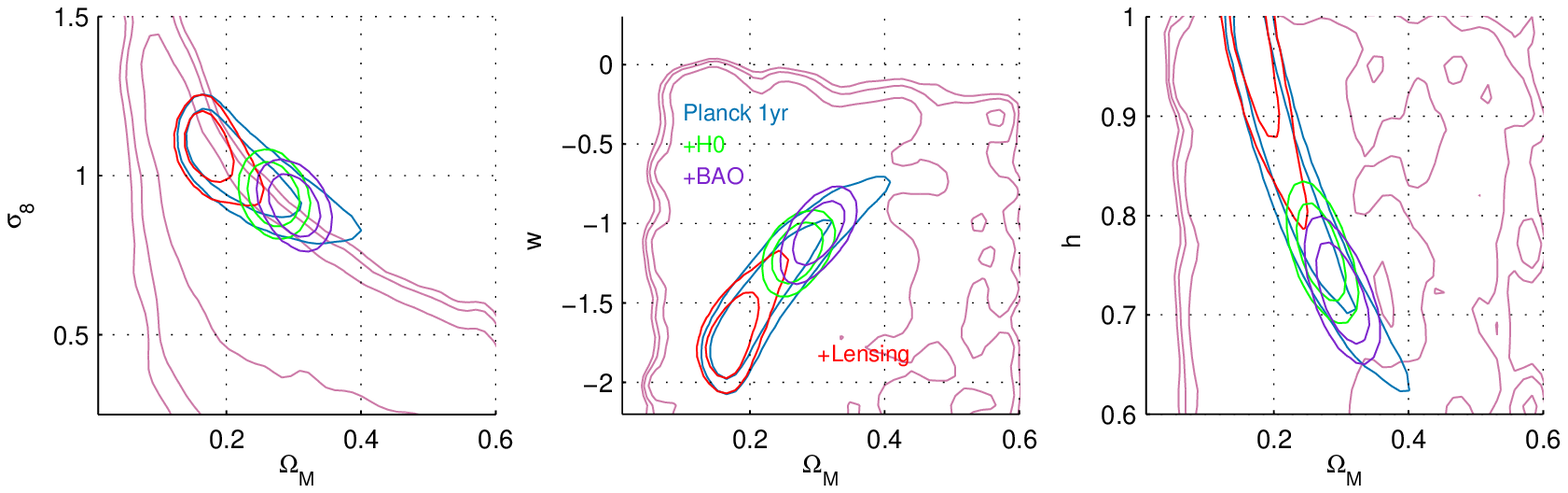, width=1.8\columnwidth, angle=0}
{\bf $k_{\rm max}\leq 5.0h$Mpc$^{-1}$ with high-k Ansatz $E_S=1$}
\psfig{file=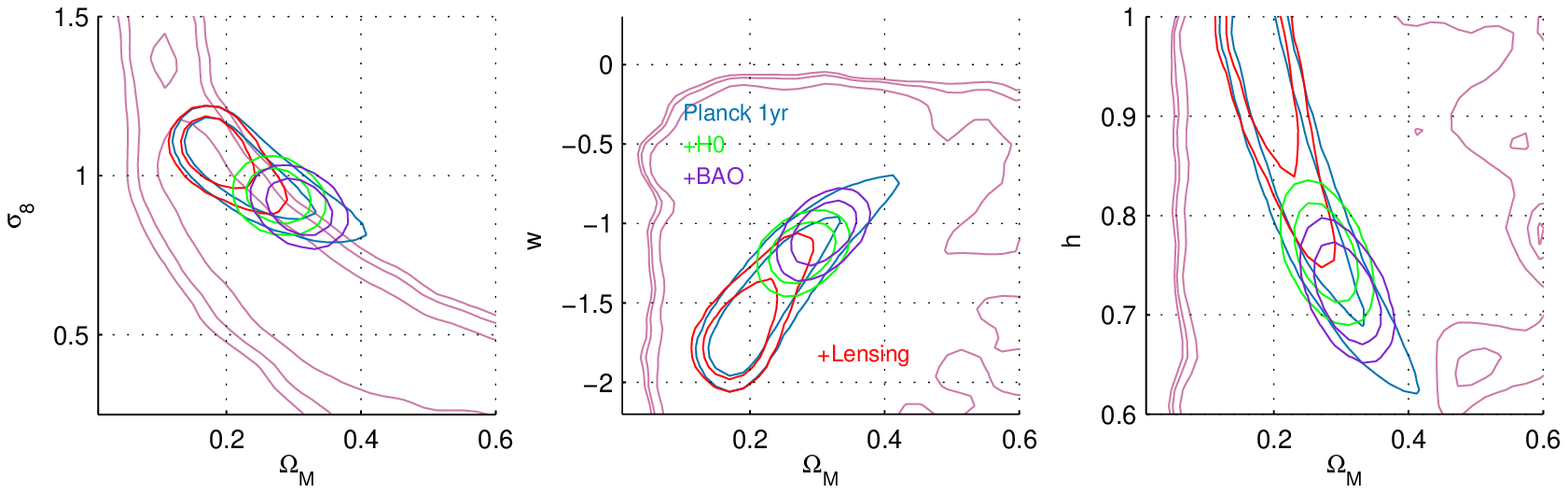, width=1.8\columnwidth, angle=0}
{\bf $k_{\rm max}\leq 5.0h$Mpc$^{-1}$ with high-k Ansatz $E_S=3$}
\psfig{file=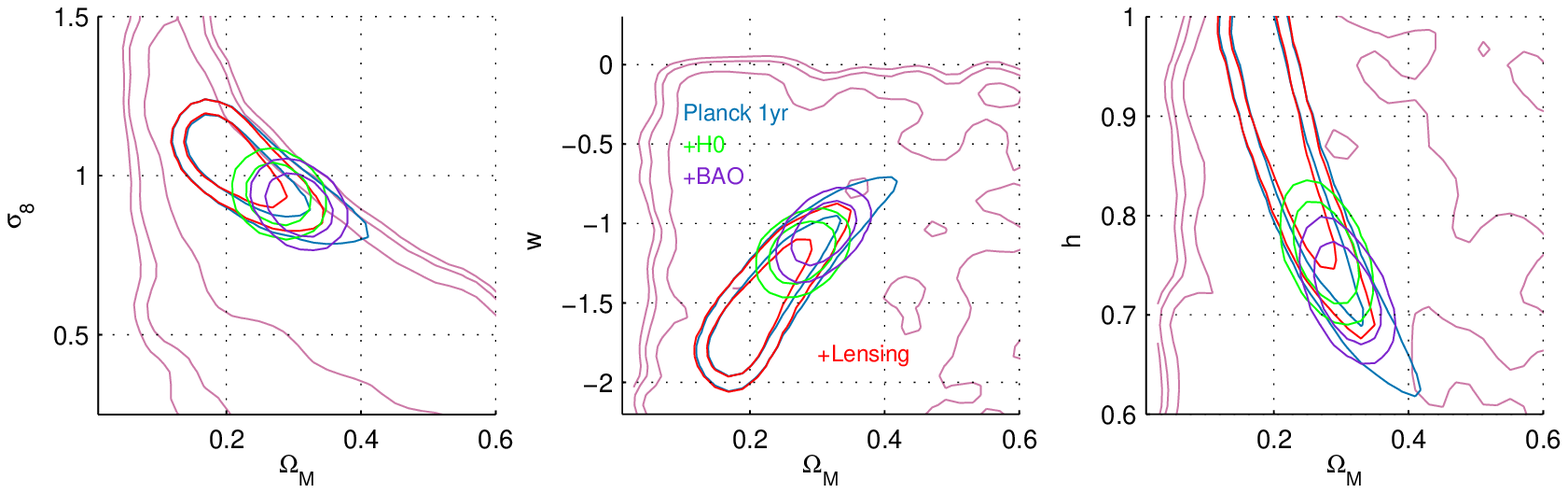, width=1.8\columnwidth, angle=0}
\caption{The combination of \emph{Planck} CMB data with the 3D cosmic shear constraints (Lensing; red contours), compared to the combination of \emph{Planck} 
with BAO (Percival et al., 2010; purple contours) and with $H_0$ (Riess et al., 2011; green contours). We show
three projected 2-parameter spaces in a wCDM cosmology, where
$w_a=0$. Contours shown are 2-parameter $1$ and $2\sigma$ confidence regions, pink contours show the lensing-only constraints. 
Note that the absence of power suppression in the range $1.5 < k \leq 5.0 h$Mpc$^{-1}$, such as 
may be provided by AGN feedback results in the posterior being driven to $\Omega_{\rm M}\ls 0.2$, $w\ls 1.5$, $h\gs 0.8$ and $\sigma_8\gs 1.0$ for the 
$k<5.0h$Mpc$^{-1}$ results. The lower two rows include the high-$k$ functional ansatz discussed in Section \ref{Scales}.}
\label{many}
\end{figure*}

\begin{figure*}
{\bf $k_{\rm max}\leq 1.5h$Mpc$^{-1}$}
\psfig{file=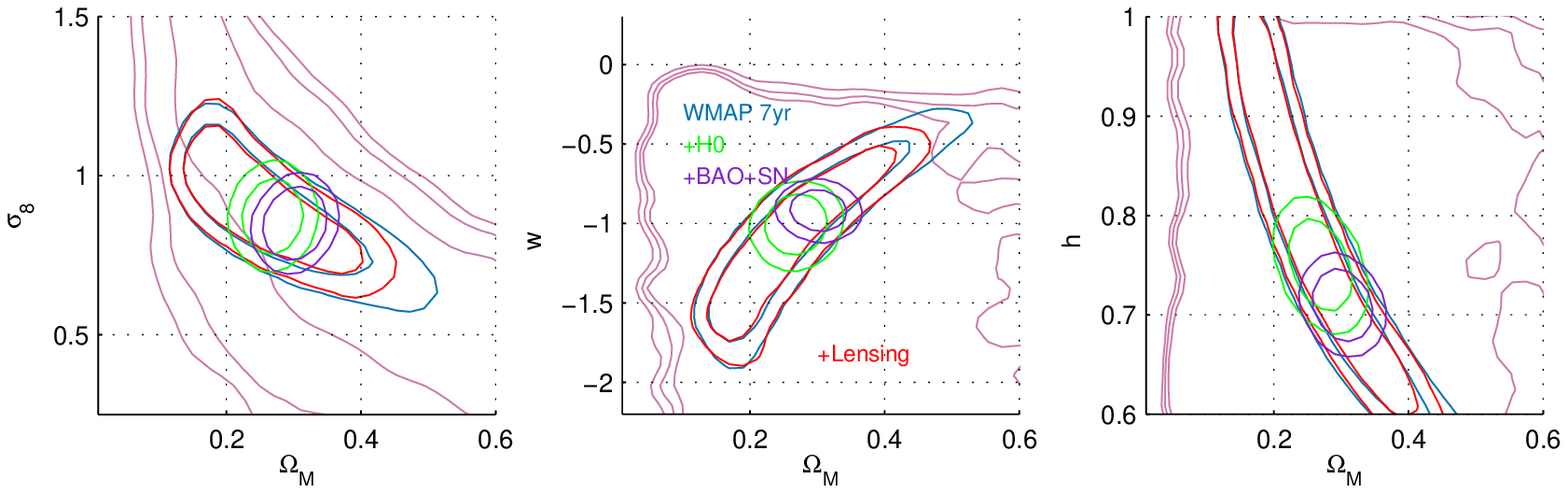, width=1.8\columnwidth, angle=0}
{\bf $k_{\rm max}\leq 5.0h$Mpc$^{-1}$}
\psfig{file=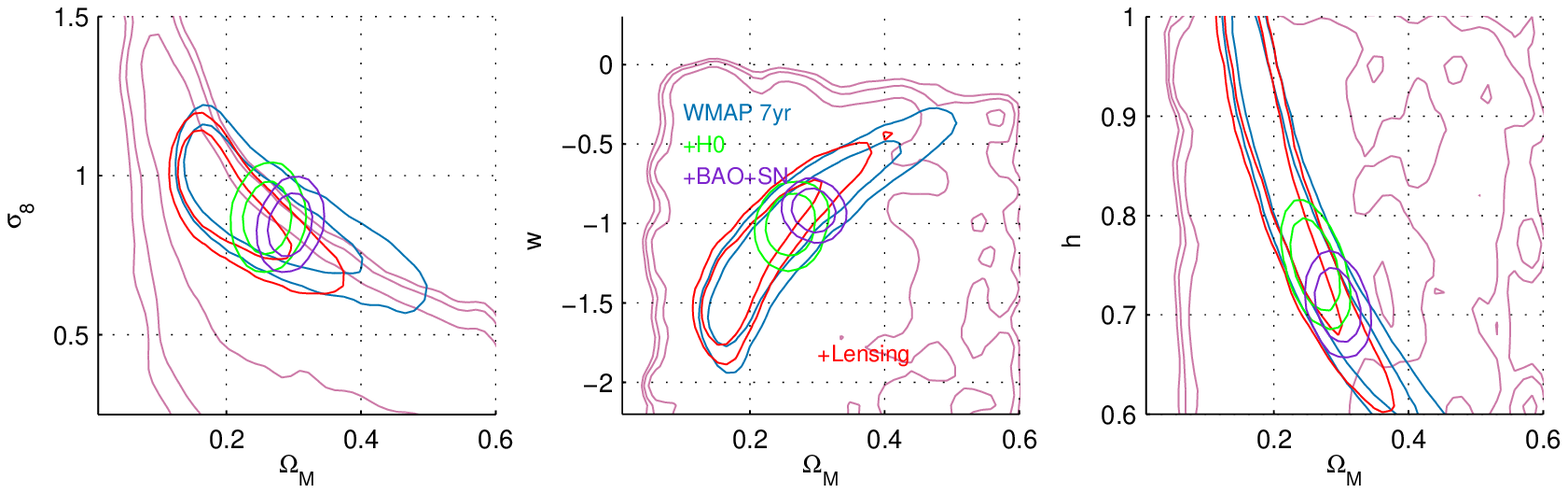, width=1.8\columnwidth, angle=0}
\caption{As the top two rows in Figure \ref{many} but in combination with WMAP7 priors.}
\label{2many}
\end{figure*}

Figures \ref{many}, \ref{2many} and Table \ref{comparison} clearly show that lensing can provide an independent 
way to lift CMB degeneracies, to a degree comparable with current $H_0$ constraints 
and the combination of BAO+SN, in particular for $\sigma_8$ and 
the dark energy equation of state $w$; but this depends on the range of scales used. 
Using scales of $k\leq 1.5 h$Mpc$^{-1}$ only we find that the lensing data does not add any significant constraining power to the 
CMB data. However, when using scales of  $k\leq 5 h$Mpc$^{-1}$, and no baryonic feedback correction, 
the tension between the slightly lower $\sigma_8$ and $\Omega_{\rm M}$ cause the 
CMB degeneracies to be lifted, but the posterior is driven to very low $\Omega_{\rm M}\ls 0.2$, high $h\gs 0.8$ and high $\sigma_8\gs 1$. 

This is evidence of the modelling of the non-linear scales being in tension with the modelling of linear 
scales, or the presence of an undetected scale-dependent systematic effect. 
The modelling of the non-linear clustering, either dark matter or baryonic feedback are possible sources of 
plausibly incorrect modelling (see Section \ref{Scales}). Alternatively a cosmological model assuming  
$w=-1$ is not correct or needs an additional component: one possible assumption that could be relaxed is 
that of no massive neutrino species, which could cause a suppression of power at scales $>1.5 h$Mpc$^{-1}$ (see Jimenez et al., 2010). 
This result is unlikely to be caused by residual intrinsic alignment contamination,
because such an effect is expected to impact all scales, but this is a further possibility. At the current time 
the data, and modelling of the baryonic feedback, are not sufficient to confidently distinguish these possibilities; although one, or more, 
of these must be causing the observed effect. As shown in Figure \ref{many}, 
we find that when the high-$k$ functional ansatz described in Section \ref{Scales} is included that the lensing 
constraints are more consistent with the \emph{Planck} constraints, and that the degeneracy lifting is relaxed. 

\subsection{LCDM \& waCDM cosmologies}
\label{LCDMwaCDM cosmologies}
\begin{figure*}
\psfig{file=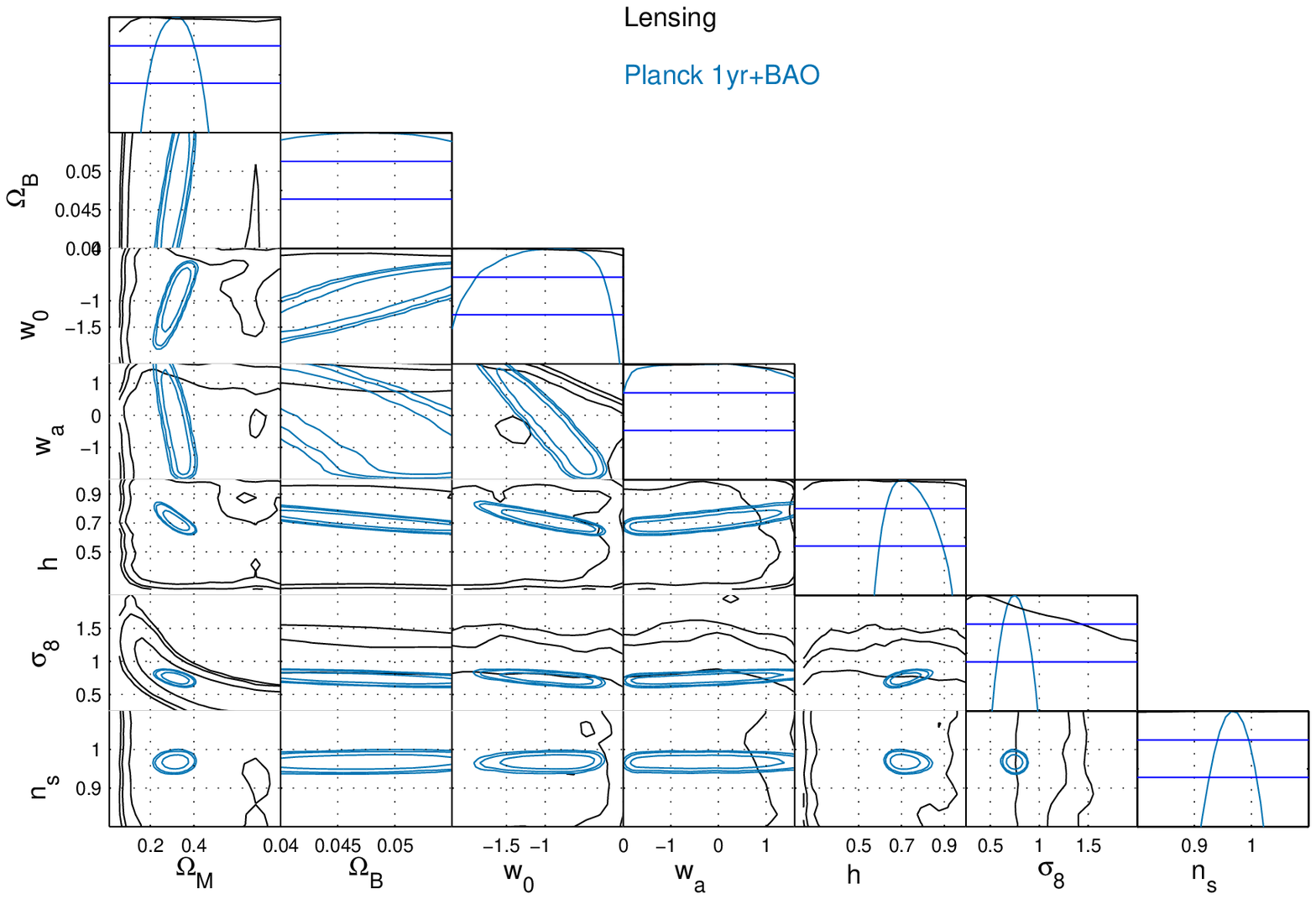, width=1.8\columnwidth, angle=0}
\caption{The cosmological parameter constraints from 3D cosmic shear in the waCDM parameter space 
with $k_{\max}=5.0h$Mpc$^{-1}$, with no baryonic feedback model included. We show each projected 2-parameter 
combination accessible in this analysis, with the 2-parameter $1$, $2$ and $3\sigma$ confidence regions shown. 
Shown are lensing (3D cosmic shear) (black) and Planck+BAO constraints (blue; for waCDM respectively). 
We also show the projected 1-parameter likelihoods for each parameter (the top-most 
blue and black lines). See Section \ref{LCDMwaCDM cosmologies} for a discussion of this Figure.}
\label{2DT}
\end{figure*}

In Figure \ref{2DT} we show the 2-parameter projected constraints for the waCDM set with a $k_{\rm max}=5.0h$Mpc$^{-1}$, 
in each of the 2-parameter combinations that are accessible in this analyses\footnote{We note that 
the predictions of Kitching (2007) are consistent with the constraints presented here -- although the realised survey geometry, number density and depth were not considered explicitly.}. 
It is clear from Figure \ref{2DT} that lensing is providing constraints consistent with 
Planck+BAO for waCDM cosmologies, but that there is very little gain over these, even at the $1\sigma$ level. 
In the waCDM parameter space the constraints from lensing are very broad as the data 
is not sufficient to place tight constraints in such a larger parameter space. 

In the LCDM parameter space the CMB alone already constrains most parameters very tightly -- 
the significant geometric degeneracy in the CMB being lifted by the choice of a cosmological parameter set that assumes flatness -- and so similarly we find that 
there is no tension with the \emph{Planck} results, but also no improvement with the addition of the CFHTLenS constraints. For comparison we find that for an 
LCDM cosmology $\sigma_8(\Omega_{\rm M}/0.27)^{0.69\pm 0.22}=1.16\pm 0.27$ compared to \emph{Planck} who find $\sigma_8(\Omega_{\rm M}/0.27)^{0.46}=0.89\pm 0.03$ using the same cosmology. 

\subsection{Comparison with 2D correlation function analyses}
Comparing these constraints with those from 2D and tomographic correlation function analysis of the CFHTLenS data 
(Kilbinger et al., 2013; Simpson et al., 2013; Benjamin et al., 2013; Heymans et al., 2013) 
we find similar constraints from the full 3D analysis on some parameters, for example $w$ ($+1.30-0.82$ in this paper 
compared to approximately $\pm 1.0$ in Kilbinger et al., 2013 for lensing alone), despite the
fact that we only probe $5\%$ to $16\%$ of the modes in the matter power spectrum, and $\sim 20\%$ fewer galaxies:
we use $k\leq 1.5h$Mpc$^{-1}$ or $5.0h$Mpc$^{-1}$ compared to $k\ls 30h$Mpc$^{-1}$ 
(see Section \ref{Scales}), however this difference results in some subtlety in the comparison that we describe here. 

In Table \ref{comparison} we show constraints on the empirical 
relation $\sigma_8(\Omega_{\rm M}/0.27)^{\alpha}=$ constant, commonly used to parameterise the amplitude and width of the
contours in the ($\sigma_8$, $\Omega_{\rm M}$) plane. 
We find weaker constraints in the orthogonal direction parameterised by $\alpha$ 
(width of the contour, as can be seen by comparing
Figure \ref{2DT} of this paper with Figures 11 and 5 of Kilbinger et al., 2013 and Heymans et al., 2013 respectively
for wCDM and LCDM cosmologies) and on $\sigma_8$. 
2D correlation functions constrain a long and thin
contour in the $(\sigma_8, \Omega_{\rm M})$ plane, and the conservative 3D cosmic shear constraints presented in this paper are wider, however 
some marginalised quantities are determined better by one method, some by the other.
One may expect simple amplitude changes in the lensing signal (such 
as changes in the orthogonal direction, or $\sigma_8$) to be measured more accurately for the correlation function 
analyses, due to the much larger number of $k$-modes analysed; and this is supported by the decrease in the error as we increase the number of $k$-modes 
in the 3D cosmic shear analysis.  
However other effects, like shape changes in the linear part of the power spectrum
(determined by combinations such as $\Omega_{\rm M}h^2$) and parameters that change the redshift evolution of the 
matter power spectrum or the expansion history such as $w$, are more well resolved by 3D cosmic shear at all redshifts. 
Hence comparable constraints are expected on these parameters, and in combination with CMB constraints,
even with a smaller number of total $k$-modes. 
We find similar constraints on $w$ comparing 2D correlation function analyses and 3D cosmic shear indicating that the extra
constraining power from including small angular scales in the 2D analysis compensates for the constraining power lost 
by not analysing the data fully in 3D. 
The tighter constraints on $\Omega_{\rm M}$, $h$, $w$ also help, through lifting 
degeneracies, in measuring other parameters in combination with CMB constraints for example $\sigma_8$.
We also find a higher value of $\sigma_8$ than the correlation function analyses,
although results are consistent at the $\sim 1\sigma$ level. In these comparisons the scale-dependence of the power, and the modelling 
uncertainties at high $k$ values, described in Section \ref{Scales} should be considered.
\begin{table*}
\begin{tabular}{|c|l@{ $\pm$ }l|l|l|}
Parameter& \multicolumn{2}{l|}{flat LCDM Lensing only}& Analysis and Method&\\
\hline
\hline
$\alpha$        &\multicolumn{2}{l|}{$0.44^{+0.24}_{-0.36}$}&3D cosmic shear power spectra& $k\leq 1.5h$Mpc$^{-1}$, no early-type galaxies\\
                &\multicolumn{2}{l|}{$0.46^{+0.37}_{-0.36}$}&3D cosmic shear power spectra$^*$& $k\leq 5.0h$Mpc$^{-1}$, no early-type galaxies\\
                &$0.59$&$0.02$&1-bin correlation function$^{\dagger}$& $k\ls 30h$Mpc$^{-1}$, all galaxies\\
                &$0.46$&$0.02$&6-bin correlation function$^{\ddagger}$& $A$ marginalised, $k\ls 30h$Mpc$^{-1}$, all galaxies\\
\hline
$\sigma_8(\Omega_{\rm M}/0.27)^{\alpha}$ &\multicolumn{2}{l|}{$1.16^{+0.27}_{-0.27}$}&3D cosmic shear power spectra&$k\leq 1.5h$Mpc$^{-1}$, no early-type galaxies\\
                                         &\multicolumn{2}{l|}{$0.69^{+0.22}_{-0.22}$}&3D cosmic shear power spectra$^*$&$k\leq 5.0h$Mpc$^{-1}$, no early-type galaxies\\
                                         &$0.79$&$0.04$&1-bin correlation function$^{\dagger}$&$k\ls 30h$Mpc$^{-1}$, all galaxies\\
                                         &\multicolumn{2}{l|}{$0.77^{+0.03}_{-0.04}$}&6-bin correlation function$^{\ddagger}$&$A$ marginalised, $k\ls 30h$Mpc$^{-1}$, all galaxies\\
\hline
\\
& \multicolumn{2}{l|}{flat wLCDM Lensing only}&&\\
\hline
\hline
$\alpha$        &\multicolumn{2}{l|}{$0.46^{+0.23}_{-0.26}$}&3D cosmic shear power spectra& $k\leq 1.5h$Mpc$^{-1}$, no early-type galaxies\\
                &\multicolumn{2}{l|}{$0.39^{+0.50}_{-0.29}$}&3D cosmic shear power spectra$^*$& $k\leq 5.0h$Mpc$^{-1}$, no early-type galaxies\\
                &$0.59$&$0.03$&1-bin correlation function$^{\dagger}$& $k\ls 30h$Mpc$^{-1}$, all galaxies\\
\hline 
$\sigma_8(\Omega_{\rm M}/0.27)^{\alpha}$ &\multicolumn{2}{l|}{$1.14^{+0.26}_{-0.30}$}&3D cosmic shear power spectra&$k\leq 1.5h$Mpc$^{-1}$, no early-type galaxies\\
                                         &\multicolumn{2}{l|}{$0.72^{+0.30}_{-0.30}$}&3D cosmic shear power spectra$^*$&$k\leq 5.0h$Mpc$^{-1}$, no early-type galaxies\\
                                         &$0.79$&$0.07$&1-bin correlation function$^{\dagger}$&$k\ls 30h$Mpc$^{-1}$, all galaxies\\
\hline
$w$ &\multicolumn{2}{l|}{$-1.40^{+1.30}_{-0.82}$}&3D cosmic shear power spectra&$k\leq 1.5h$Mpc$^{-1}$, no early-type galaxies\\
    &\multicolumn{2}{l|}{$-1.41^{+1.25}_{-0.80}$}&3D cosmic shear power spectra$^*$&$k\leq 5.0h$Mpc$^{-1}$, no early-type galaxies\\
    &\multicolumn{2}{l|}{$-1.17^{+0.80}_{-1.40}$}&1-bin correlation function$^{\dagger}$&$k\ls 30h$Mpc$^{-1}$, all galaxies\\
\hline
\\
  & \multicolumn{2}{l|}{flat wCDM Lensing+WMAP7}& &\\
\hline
\hline
$\Omega_{\rm M}$& $0.252$&$0.079$&3D cosmic shear power spectra&$k\leq 1.5h$Mpc$^{-1}$, no early-type galaxies\\
                & $0.210$&$0.069$&3D cosmic shear power spectra$^*$&$k\leq 5.0h$Mpc$^{-1}$, no early-type galaxies\\
                & $0.325$&$0.082$&1-bin correlation function$^{\dagger}$& $k\ls 30h$Mpc$^{-1}$, all galaxies\\
                & $0.256$&$0.110$&6-bin correlation function$^{\ddagger}$&$A$ marginalised, $k\ls 30h$Mpc$^{-1}$, all galaxies\\
\hline
$\sigma_8$      & $0.88$&$0.23$&3D cosmic shear power spectra&$k\leq 1.5h$Mpc$^{-1}$, no early-type galaxies\\
                & $0.88$&$0.22$&3D cosmic shear power spectra$^*$&$k\leq 5.0h$Mpc$^{-1}$, no early-type galaxies\\
                & $0.77$&$0.11$&1-bin correlation function$^{\dagger}$& $k\ls 30h$Mpc$^{-1}$, all galaxies\\
                & $0.81$&$0.10$&6-bin correlation function$^{\ddagger}$&$A$ marginalised, $k\ls 30h$Mpc$^{-1}$, all galaxies\\
\hline
$w$             & $-1.16$&$0.38$&3D cosmic shear power spectra&$k\leq 1.5h$Mpc$^{-1}$, no early-type galaxies\\
                & $-1.23$&$0.34$&3D cosmic shear power spectra$^*$&$k\leq 5.0h$Mpc$^{-1}$, no early-type galaxies\\
                & $-0.86$&$0.22$&1-bin correlation function$^{\dagger}$& $k\ls 30h$Mpc$^{-1}$, all galaxies\\
                & $-1.05$&$0.34$&6-bin correlation function$^{\ddagger}$&$A$ marginalised, $k\ls 30h$Mpc$^{-1}$, all galaxies\\
\hline
$h$             & $0.78$&$0.12$&3D cosmic shear power spectra&$k\leq 1.5h$Mpc$^{-1}$, no early-type galaxies\\
                & $0.83$&$0.12$&3D cosmic shear power spectra$^*$&$k\leq 5.0h$Mpc$^{-1}$, no early-type galaxies\\
                & $0.66$&$0.11$&1-bin correlation function$^{\dagger}$& $k\ls 30h$Mpc$^{-1}$, all galaxies\\
                & $0.74$&$0.14$&6-bin correlation function$^{\ddagger}$&$A$ marginalised, $k\ls 30h$Mpc$^{-1}$, all galaxies\\
\hline
\hline
\end{tabular}
\caption{The mean parameter values from an CFHTLenS 3D cosmic shear analysis (this paper) and we quote 
numbers from 1-bin 2D correlation function 
analysis ($^{\dagger}$Kilbinger et al., 2013, Tables 2 and 3) and 6-bin tomographic correlation 
function analysis ($^{\ddagger}$Heymans et al, 2013, Tables 2 and 3). 
The upper rows show the lensing-only constraints on the empirical relation $\sigma_8(\Omega_{\rm M}/0.27)^{\alpha}=$constant, that 
parameterises the amplitude and width of the ($\sigma_8$, $\Omega_{\rm M}$) contour; see Heymans et al. (2013) Table 2 for 
further values of these under various assumptions in a $\Lambda$CDM cosmology. 
The lower rows compare the constraints in a flat wCDM cosmology, with lensing combined with WMAP7. 
For the 3D cosmic shear only constraints we quote asymmetric error bars. 
Note that Heymans et al., (2013) also marginalised over an IA parameter $A$, but included $1.5$ times as many galaxies. 
The errors are symmetric 1-parameter $1\sigma$ values. $^*$Have no allowance for baryonic feedback effects that are likely to impact 
constraints using $k\gs 1.5h$Mpc$^{-1}$. }
\label{comparison}
\end{table*}

Finally we note here a further aspect that may contribute to the differences with 2D and tomographic analyses. 
The approach of using the analytic covariance (and 
one-point estimator) in this paper does not suffer from noise due to finite number of simulations: the $1656$ 
independent lines of sight in the CFHTLenS {\sc Clone} (Harnois-Deraps et al., 2012) 
and the $210$ data points used in Heymans et al. (2013) leads to a $\simeq 4$ per cent fractional error on the 
inverse covariance in their study 
(using the scaling of Taylor, Joachimi, Kitching, 2013), an error that is not present in the 3D cosmic shear analysis presented here; 
this error could be considered as a lower bound since any other sources of error or bias in the simulations would not 
be captured in this number.

A more quantitative comparison of 2D correlation function methodology and 3D power spectrum analysis 
is more complicated and beyond the scope of this paper; our aim is to present the 3D cosmic shear 
results, not to perform a close and comprehensive 2D correlation function to 3D power spectrum comparison. 
In particular for a close comparison with Kilbinger et al. (2013) there are several differences 
in their analysis that would have to be considered. For example a different galaxy selection was made, different ranges in scale were used. 

\subsection{Comparison with expected constraints}
\begin{figure*}
\psfig{file=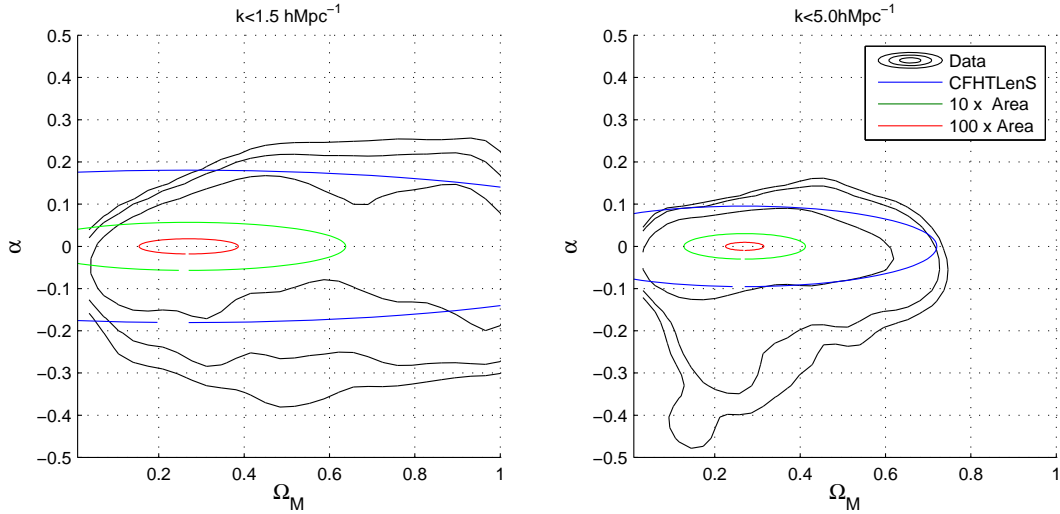, width=1.7\columnwidth, angle=0}
\caption{The constraints from CFHTLenS 3D cosmic shear only in the $(\alpha, \Omega_{\rm M})$ plane marginalised over the other parameters in the wCDM
parameter set. This is the plane in which the constraints in
the projected $(\sigma_8,\Omega_{\rm M})$ plane are uncorrelated, which is parameterised by the function $\sigma_8(\Omega_{\rm M}/0.27)^{\alpha}=$constant.
The black contours show the $1$, $2$, $3\sigma$ constraints from the data, the blue lines show the expected $1\sigma$ two-parameter projected
constraints in this plane from a Fisher matrix analysis for CFHTLenS, the green and red lines show the expected constraints for a survey $10$ times larger
area and $100$ times larger in area respectively with all other survey characteristics kept constant.}
\label{alpha}
\end{figure*}

The constraints from lensing alone presented in this paper are conservative in that we remove galaxies that have evidence for an intrinsic alignment signal, and 
we remove scales for which there exist uncertainties in the non-linear modelling of the matter power spectrum. However, we can predict the expected constraints 
using such conservative assumptions and galaxy selections using the Fisher matrix formalism (Kitching, Heavens, Miller, 2013; which is also used as the proposal
distribution for the MCMC chains in this paper). We find that the most tightly constrained plane is in the projected $(\sigma_8$, $\Omega_{\rm M})$ direction; see 
Figure \ref{2DT}, that can be parameterised by $\sigma_8(\Omega_{\rm M}/0.27)^{\alpha}=$constant. 
In Figure \ref{alpha} we show the measured and predicted constraints on the parameters $\alpha$ and $\Omega_{\rm M}$ marginalising over all other parameters in 
the wCDM parameter set. 

We find that the expected $1\sigma$ contours coincide with the measured constraints. Therefore the analysis, whilst conservative in 
its assumptions which leads to broad constraints, is as expected for a survey the size of CFHTLenS. We also show predictions for surveys that are $10$ and $100$ 
times the area of CFHTLenS, keeping all other survey characteristics the same, for which we expect significant gains.  

\section{Conclusions}
\label{Conclusion}
In this paper we present the first application of the 3D cosmic shear method to a wide-field weak lensing 
survey, CFHTLenS (Erben et al., 2013; Heymans et al., 2012) and use this method to measure cosmological parameters 
including the dark energy equation of state parameters $w_0$ and $w_a$. The CFHTLenS data covers $154$ 
square degrees, of which $61\%$ is unmasked and passes systematics tests, and has been analysed with the 
state-of-art in shape measurement (\emph{lens}fit, Miller et al., 2013) and redshift estimation ({\sc BPZ}, 
Hildebrandt et al., 2012). 

3D cosmic shear, which uses the covariance of the 3D spherical harmonic/spherical Bessel coefficients of the shear field as its signal, 
has a number of useful features over other approaches in that i) it does not bin the data, in particular in the redshift direction 
along which discoveries of redshift-dependent effects may be found (i.e. dark energy) ii) it allows for a control of the angular ($\ell$) 
and radial ($k$) modes in the analysis independently which means that non-linear modes in the matter power spectrum may be explicitly 
excluded iii) it allows for extra information from individual galaxies to be used, 
for example the posterior information in 
redshift and iv) it uses a one-point estimator with an analytic covariance estimate, and hence in this analysis is not sensitive to estimating the inverse covariance from simulations.  
To account for angular masks in the analysis we present a pseudo-$C_{\ell}$ method for 3D fields and apply this to the analytic covariance. This is the first 
application of a pseudo-$C_{\ell}$ method on weak lensing data: previously mask window functions 
have been computed for the galaxy power spectrum in Pen et al. (2003) and were not taken into account in Brown et al. (2003), Heymans et al. (2005) or Kitching et al. (2007).  

One can project the shear field onto 2D planes in redshift, to create tomographic slices, and in this paper 
(and in Kitching, Heavens, Miller, 2011) we show how 2D and tomographic 
power spectra can be recovered from the full 3D shear field. Here we apply this and present 2D and tomographic cosmic shear $C(\ell)$ power spectra (the first presentation of a tomographic cosmic shear power spectrum from data). 
To reduce intrinsic alignment systematics we exclude all galaxies with {\sc BPZ} parameter $T_B\leq 2$ (this preferentially selects late-type galaxies). In the future mitigation techniques should be developed to either remove or account for intrinsic alignments. 
3D cosmic shear uses the redshift posterior probabilities for each galaxy $p_g(z)$ in the estimator. We justify our use of the $p_g(z)$ on 
the analysis of Benjamin et al. (2012) who found that the redshift probability distributions were unbiased with respect to a spectroscopic 
sample of the same galaxies, however that analysis used a correlation function technique, in only 6 redshift bins and for all galaxies in the CFHTLenS catalogue. This means
that it probed much smaller scales than in this paper, averaged over all galaxy types, and used a much coarser redshift sampling.  
In future the determination of the fidelity of redshift
posterior information should be performed over scales, and with a redshift sampling, 
matched to those used in any 3D cosmic shear analysis. 

The results we find are not formally in conflict with previous correlation function (configuration space) analyses of CFHTLenS (Kilbinger et al., 2013; Benjamin 
et al., 2013; Simpson et al., 2013; Heymans et al. 2013). The most interesting finding we have comes from the ability of 3D cosmic shear to probe rather well defined 
ranges of physical wavenumber. 
For wCDM cosmologies we find evidence that there is tension between the lensing 
constraints that only use scales $k\leq 1.5h$Mpc$^{-1}$ and those that use $k\leq 5h$Mpc$^{-1}$, where the lensing at smaller scales (higher $k$) prefers a lower value 
of $\sigma_8$. 
Taken at face value in combination with either \emph{Planck} or WMAP7 CMB priors the smaller scales lift the CMB degeneracy and favour $w\ls -1.5$, 
a high value of $h\gs 0.8$, a low $\Omega_{\rm M}\ls 0.2$ and a high $\sigma_8\gs 1.0$ in comparison to concordance LCDM values.
This is evidence of either the non-linear modelling being in tension with the linear model, and/or the cosmological model having a deviation from the 
concordance LCDM: possible explanations include 
the effects of AGN feedback on the non-linear matter power spectrum (see van Daalen et al., 2011; Semboloni et al., 2011, 2013), 
or an additional component (for example a massive neutrino species) that can suppress power at small scales. We find that when we include a functional ansatz 
that models the damping of the matter power spectrum due to AGN feedback on scales $k\gs 1.5 h$Mpc$^{-1}$, that the cosmological constraints for the two 
different ranges of scale are in better agreement. We leave a full investigation of these possible effects for future work, but 
note that one or more of these explanations is required to explain this observation. 
 

\newpage
\noindent{\em Acknowledgements:} 
We thank Dipak Munshi, Andy Taylor, Fergus Simpson, Stephen Feeney, Hiranya Peiris, Licia Verde, Raul Jimenez, Jason McEwen 
and Mark Cropper for useful discussions. We thank Anthony Lewis and the developers of {\sc camb}, and the PPF 
module for making their code public. We thank Eric Tittley and Mark Holliman for system 
administration on several machines used in this work. We made use of 
CosmoCalc \url{http://www.astro.ucla.edu/~wright/CosmoCalc.html} (Wright, 2006) 
during development. We thank WMAP for providing their MCMC chains available for download 
\url{http://lambda.gsfc.nasa.gov}, and Eiichiro Komatsu for providing supplementary data 
products to the main WMAP data release. We thank ESA \emph{Planck} for providing their MCMC chains 
available for download \url{http://pla.esac.esa.int/pla/aio/planckProducts.html}. 
TDK acknowledges support from a Royal Society University Research Fellowship.

We refer to Heymans et al. (2013), and Fu et al. (2014) for a full list of acknowledgments for CFHTLenS. 

{\small \emph Author Contributions: All authors contributed to the
  development and writing of this paper.  
  The authorship list reflects
  the lead authors of this paper (TK, AH) followed by two
  alphabetical groups.  The first alphabetical group includes key
  contributors to the science analysis and interpretation in this
  paper, the founding core team and those whose long-term significant
  effort produced the final CFHTLenS data product.  The second group
  covers members of the CFHTLenS team who made a significant
  contribution to either the project, this paper, or both.  The
  CFHTLenS collaboration was co-led by CH and LVW. 
  The CFHTLenS Cosmology Working Group was led by TK.} 



\onecolumn
\section*{Appendix A: E and B-mode Separation} 
In Kitching et al. (2007) there was no attempt made to split the shear
estimators into E and B-mode parts, in analogy to the procedure that is
followed for correlation function analysis. We show here how to
correctly link the raw calculated spherical harmonic coefficients into
those that can be used to compute E and B-mode power. 

We start with the standard relation between shear and the Newtonian potential
$\Phi$ (see Castro et al., 2003 for notation)
\be
\bgamma(\btheta)=\frac{1}{2}\partial\partial\Phi(\btheta)
\ee
where in Fourier space the complex derivative $\partial=\partial_x+{\rm i}\partial_y$ 
can be written like $(\ell_y^2-\ell_x^2)-2\idot\ell_x\ell_y$ (i.e. taking the complex
derivative of ${\rm e}^{\idot \ell.\btheta}$). We can decompose
the Newtonian potential into an E-mode part and a systematic B-mode
part in Fourier space $\phi_E+\idot\phi_B$, where $\phi$ is the
Fourier transform of $\Phi$. Therefore in Fourier space the relation
between shear and the Newtonian potential is 
\be
\bgamma(k,\ell)=(D_1+\idot D_2)(\phi_E+\idot\phi_B)
\ee
where $D_1=\frac{1}{2}(\ell_y^2-\ell_x^2)$ and
$D_2=-\ell_x\ell_y$. These can be expanded to give 
\be
\label{gD}
\bgamma(k,\ell)={\mathbb R}[\gamma(k,\ell)]+\idot{\mathbb I}[\gamma(k,\ell)]=(D_1\phi_E-D_2\phi_B)+\idot
(D_1\phi_B+D_2\phi_E), 
\ee
where on the lefthand side we have a data vector, with real and
imaginary parts, and on the righthand side we have theory.

From the data we have four vectors  
of shear, two real and two imaginary. Neglecting the weighting
functions (which do not affect this result) these are 
\be 
\label{gT}
\gamma(k,\ell)=\{{\mathbb R}[\gamma_1(k,\ell)]+\idot{\mathbb I}[\gamma_1(k,\ell)]\}+\idot\{{\mathbb R}[\gamma_2(k,\ell)]+\idot{\mathbb I}[\gamma_2(k,\ell)]\}
\ee
where $\gamma_1$ and $\gamma_2$ (defined as the shear $\bgamma=\gamma_1+\idot\gamma_2$ inferred from an observed 
ellipticity $e_{\rm obs}=(a-b)/(a+b){\rm exp}(-2\idot\theta)$ where $a$, $b$ and $\theta$ are the semimajor, semiminor axes and orientation; 
see Miller et al, 2013) have real and imaginary parts
respectively. These four components can be written as 
\ba 
\bgamma(k,\ell)&=&\sum_g (e_{1,g}+\idot e_{2,g})j_{\ell}(k r[z^f_{g}]){\rm e}^{-\idot \ell.\btheta_g}=\sum_g (e_{1,g}+\idot e_{2,g})j_{\ell}(k r[z^f_{g}])[\cos(\ell.\btheta_g)-\idot\sin(\ell.\btheta_g)]
\ea
which gives  
\ba
{\mathbb R}[\gamma_1(k,\ell)]&=& \sum_g e_{1,g}j_{\ell}(k r[z^f_{g}])\cos(\ell.\btheta_g)\nn
-{\mathbb I}[\gamma_1(k,\ell)]&=&\sum_g e_{1,g}j_{\ell}(k r[z^f_{g}])\sin(\ell.\btheta_g)\nn
{\mathbb R}[\gamma_2(k,\ell)]&=& \sum_g e_{2,g}j_{\ell}(k r[z^f_{g}])\cos(\ell.\btheta_g)\nn
-{\mathbb I}[\gamma_2(k,\ell)]&=&\sum_g e_{2,g}j_{\ell}(k r[z^f_{g}])\sin(\ell.\btheta_g),
\ea
where ${\mathbb R}$ and ${\mathbb I}$ mean real and imaginary parts respectively.

What we want is a shear estimator for E and B
modes 
\ba
\label{obsshear}
\gamma_E(k,\ell)&=&(D_1+\idot D_2)\phi_E\nn
\gamma_B(k,\ell)&=&(D_1+\idot D_2)\phi_B.
\ea
The question we address here is how to combine ${\mathbb R}[\gamma(k,\ell)]$ and
${\mathbb I}[\gamma(k,\ell)]$ to generate what
we require. Note that from equations (\ref{gD}) and (\ref{gT}) we have 
\ba 
\label{obscoeff}
{\mathbb R}[\gamma(k,\ell)]&=&(D_1\phi_E-D_2\phi_B)={\mathbb R}[\gamma_1(k,\ell)]-{\mathbb I}[\gamma_2(k,\ell)]\nn
{\mathbb I}[\gamma(k,\ell)]&=&(D_1\phi_B+D_2\phi_E)={\mathbb I}[\gamma_1(k,\ell)]+{\mathbb R}[\gamma_2(k,\ell)],
\ea
so it is tempting to associate directly $\gamma_1$ and $\gamma_2$
to the respectively signed parts
(e.g. $D_1\phi_E\equiv{\mathbb R}[\gamma_1(k,\ell)]$ etc.), 
however this would not be correct because it would neglect E and B-mode power resulting
from a mixtures of $\gamma_1$ and $\gamma_2$. Rearranging equation (\ref{gD}) gives 
\ba
\phi_E&=&\frac{1}{D_1^2+D_2^2}(D_1{\mathbb R}[\gamma(k,\ell)]+D_2{\mathbb I}[\gamma(k,\ell)])\nn
\phi_B&=&\frac{1}{D_1^2+D_2^2}(D_1{\mathbb I}[\gamma(k,\ell)]-D_2{\mathbb R}[\gamma(k,\ell)])
\ea
so that we find 
\ba 
\label{eb}
\gamma_E(k,\ell)&=&\frac{D_1}{D_1^2+D_2^2}\{D_1{\mathbb R}[\gamma(k,\ell)]+D_2{\mathbb I}[\gamma(k,\ell)]\}+\idot\frac{D_2}{D_1^2+D_2^2}\{D_1{\mathbb R}[\gamma(k,\ell)]+D_2{\mathbb I}[\gamma(k,\ell)]\}\nn
\gamma_B(k,\ell)&=&\frac{D_1}{D_1^2+D_2^2}\{D_1{\mathbb I}[\gamma(k,\ell)]-D_2{\mathbb R}[\gamma(k,\ell)]\}+\idot\frac{D_2}{D_1^2+D_2^2}\{D_1{\mathbb I}[\gamma(k,\ell)]-D_2{\mathbb R}[\gamma(k,\ell)]\}.
\ea
This now links the raw calculated spherical harmonics to an E and
B-mode representation. 

\newpage
\section*{Appendix B: The impact of shape measurement bias}
The shape measurement correction found by Heymans et al. (2013) (that needs to be applied to the data vector after the measurement process) 
is a scalar function, and is defined for each galaxy individually, $m_g$. Therefore it
acts in a similar way as a position and redshift dependent weight map or mask. Here we show how to construct unbiased 3D shear coefficients using such a function.

In a similar way to equation (\ref{a}) we take the transform of $(1+m_g)$ as
\be
m(k,\ell)=\sqrt{\frac{2}{\pi}}\sum_{g} (1+m_g) j_{\ell}(kr^0_g){\rm e}^{-\idot\bell.\btheta_g}W(r^0_g)=m_R(k,\ell)+\idot m_I(k,\ell),
\ee
where we explicitly label the real and imaginary parts of the transform $m_R$ and $m_I$. The quantity $m_g$ 
is a function of signal-to-noise and galaxy size,
and therefore may change as a function of position and redshift. As a result the coefficients $m(k,\ell)$ may have structure in both the $k$ and $\ell$
directions. If the bias was zero for all galaxies $m_g\equiv 0$ then the transform would result in coefficients that we label with a zero $m^0_R(k,\ell)+\idot m^0_I(k,\ell)$.

We can therefore write a correct set of shear coefficients, generalising the approach of Heymans et al. (2013) to the complex transform case
\be
\label{mcorr1}
\bgamma^{\rm corrected}(k,\ell)=\left({\mathbb R}[\gamma(k,\ell)]+\idot{\mathbb I}[\gamma(k,\ell)]\right)\left(\frac{m^0_R(k,\ell)+\idot m^0_I(k,\ell)}{m_R(k,\ell)+\idot m_I(k,\ell)}\right),
\ee
using the notation from Appendix A for the real and imaginary part of the shear coefficients. This expression corrects the coefficient using the $(1+m_g)$ factor but
ensures that the original coefficients are recovered in the limit that $m_g\rightarrow 0$.

Equation (\ref{mcorr1}) can be expanded such that
\be
\label{mcorr2}
\bgamma^{\rm corrected}(k,\ell)=\left({\mathbb R}[\gamma(k,\ell)]+\idot{\mathbb I}[\gamma(k,\ell)]\right)\left(\frac{M_R+\idot M_I}{M_N}\right),
\ee
where $M_R=m_Rm^0_R+m_Im^0_I$, $M_I=m^0_Im_R-m^0_Rm_I$, $M_N=m_R^2+m_I^2$, and we have suppressed the variable $(k,\ell)$ for clarity; note that if $m_g=0$ then
the imaginary part of the correction is zero $M_I=0$ and $M_R=M_N$. Using equations (\ref{eb}) from Appendix A and substituting the above
we can now write corrected E and B-mode coefficients as
\ba
{\mathbb R}[\gamma^{\rm corrected}_E(k,\ell)]&=&\frac{M_R}{M_N}{\mathbb R}[\gamma_E(k,\ell)]-\frac{M_I}{M_N}{\mathbb R}[\gamma_B(k,\ell)]\nn
{\mathbb R}[\gamma^{\rm corrected}_B(k,\ell)]&=&\frac{M_R}{M_N}{\mathbb R}[\gamma_B(k,\ell)]+\frac{M_I}{M_N}{\mathbb R}[\gamma_E(k,\ell)]\nn
{\mathbb I}[\gamma^{\rm corrected}_E(k,\ell)]&=&\frac{M_R}{M_N}{\mathbb I}[\gamma_E(k,\ell)]-\frac{M_I}{M_N}{\mathbb I}[\gamma_B(k,\ell)]\nn
{\mathbb I}[\gamma^{\rm corrected}_B(k,\ell)]&=&\frac{M_R}{M_N}{\mathbb I}[\gamma_B(k,\ell)]+\frac{M_I}{M_N}{\mathbb I}[\gamma_E(k,\ell)].
\ea
Thus we see that the shape measurement bias mixes E and B-modes together, and that this must be corrected for in the coefficients. The corrected coefficients are used in
this paper in the likelihood analysis.

Finally, the variance of the corrected spherical harmonic coefficients can be related to the variance of the observed ellipticities by
\ba
\sigma^2_{{\mathbb R}\gamma_E}&=&\left(\frac{D_1^2}{D_1^2+D_2^2}\right)\left(\frac{M_R^2+M_I^2}{M_N^2}\right)\sigma^2_{e}\nn
\sigma^2_{{\mathbb I}\gamma_E}&=&\left(\frac{D_2^2}{D_1^2+D_2^2}\right)\left(\frac{M_R^2+M_I^2}{M_N^2}\right)\sigma^2_{e}
\ea
assuming that the variance of the $e_1$ and $e_2$ components $\sigma_e$ are equal, although this assumption can be relaxed. In Section \ref{Noise Covariance} we express
the variance above as a complex number $\sigma^2_{\epsilon}=\sigma^2_{{\mathbb R}\gamma_E}+{\rm i}\sigma^2_{{\mathbb I}\gamma_E}$.

\newpage
\section*{Appendix C: Pseudo-Estimators in 3D}
This Appendix is based on the formalism first presented in Munshi et al. (2011), we reproduce the derivation here to 
match to the notation used for the covariance and adopt a flat-sky approximation. 
We show here how a mixing matrix can be defined that, in a forward convolution with the 
analytic covariance, results in a `pseudo' covariance estimate that now accounts for angular masks in a survey.  

In 3D cosmic shear we expand in a radial wavenumber $k$ as well as an azimuthal wavenumber $\ell$. Such that in the flat-sky limit we can
relate the observed shear, as a function of radius $r$ and sky coordinate $\btheta$, to some spherical harmonic modes as
\be
\gamma(r,\btheta)=\left(\frac{2}{\pi}\right)^{1/2}\int {\rm d}k\int \frac{{\rm d}^2\bell}{(2\pi)^2}\gamma(k,\bell)j_{\ell}(kr){\rm e}^{\idot\bell.\btheta}
\ee
and its associated inverse. If we assume that the real shear field is masked by real (scalar) mask $K(\btheta)$ 
such that $\gamma(r,\btheta)\rightarrow \gamma(r,\btheta)K(\btheta)$ then the masked coefficients are given by 
\be 
\hat\gamma(k,\bell)=\left(\frac{2}{\pi}\right)^{1/2}\int {\rm d}r r^2 \int {\rm d}^2\btheta \gamma(r,\btheta)K(\btheta)W(r)j_{\ell}(kr){\rm e}^{-\idot\bell.\btheta}.
\ee
Expanding the unmasked shear field and the mask field in spherical Bessel coefficients, and integrating over angle, the 
masked coefficients can be written in a compact form as 
\be 
\label{pcoeff}
\hat\gamma(k,\bell)=\int \frac{{\rm d}^2\bell'}{(2\pi)^2}\int {\rm d}k' K(\bell-\bell')\gamma(k',\bell')F_{\ell\ell'}(k,k')
\ee
where
\be 
F_{\ell\ell'}\equiv \left(\frac{2}{\pi}\right)\int r^2{\rm d}r j_{\ell}(k'r)j_{\ell'}(kr)W(r),
\ee
where $W(r)$ is an arbitrary weight function, and $K(\bell)$ is the Fourier transform of the mask 
$K(\btheta)=\int \frac{{\rm d}^2\bell}{(2\pi)^2}K(\bell){\rm e}^{\idot\bell.\btheta}$.
Assuming the extended Limber approximation (LoVerde \& Afshordi, 2008), which is sufficient at large $\ell\gs 100$, 
we can simplify the matrix $F$ by replacing the Bessel functions with 
$j_{\ell}(kr)\approx \left(\frac{\pi}{2}\right)^{1/2}\frac{1}{\ell^{1/2}}\delta^D(kr-\ell)$. 

Taking the covariance of equation (\ref{pcoeff}) we find that 
\be 
\label{c}
\widetilde C_{\ell}(k_1,k_2)=\int {\rm d}\ell' \frac{\ell}{\ell'} \frac{1}{k_1^2k_2^2}M^{3D}_{\ell\ell'} C_{\ell'}\left(\frac{\ell'}{\ell}k_1, \frac{\ell'}{\ell}k_2\right),
\ee
where $M^{3D}_{\ell\ell'}$ is a mixing matrix 
\be
M^{3D}_{\ell\ell'}=\frac{\ell'}{(2\pi)^2}\int_0^{2\pi} {\rm d}\psi |K(L)|^2, 
\ee
and $L^2=\ell^2+\ell'^2-2\ell\ell'\cos(\psi)$. This expression then takes into account the full 2D structure of the mask. 
This is the expression we use to account for the masks, on the theory side, in the likelihood analysis presented. 

We find that the mask only mixes $\ell$-modes in the signal part of the covariance, not the shot noise part which is only affected through an area scaling. This can be 
shown following the derivation of the shot noise covariance in Kitching (2007). 

\newpage
\section*{Appendix D: 2D power from 3D power}
Here we show how 2D and tomographic cosmic shear power spectra can be calculated
from the full 3D cosmic shear power spectrum $C_{\ell}(k,k')$.

We start by defining the projected 2D spherical harmonic coefficients
as 
\be 
\bgamma^{2D}_{\ell m}(\Delta r)=\int {\rm d}\theta{\rm d}\phi
       {}_{\pm 2}Y^m_{\ell}(\theta,\phi) \int_{\Delta r} {\rm d}r \gamma(\theta,\phi,r)
       W(r),
\ee
where $W(r)$ is some arbitrary weight function; we explicitly label
the integral over $r$ with the range $\Delta r$, which is the `bin
width' of the 2D power to be calculated. Replacing
$\gamma(\theta,\phi,r)$ with its spherical harmonic transform
we have 
\be 
\bgamma^{2D}_{\ell m}(\Delta r)=\int {\rm d}\theta{\rm d}\phi
       {}_{\pm 2}Y^m_{\ell}(\theta,\phi) \int_{\Delta r} {\rm d}r \left\{\int
       \sum_{\ell'm'}\bgamma_{\ell'm'}(k)j_{\ell'}(kr)
           {}_{\pm 2}Y^{*m'}_{\ell'}(\theta,\phi) {\rm d}k \right\}
       W(r).
\ee
Using the relation between the ${}_{\pm 2}Y^m_{\ell}$ we find
that 
\be 
\bgamma^{2D}_{\ell m}(\Delta r)=\int_{\Delta r} {\rm d}r \int {\rm d}k
j_{\ell}(kr)W(r)\bgamma_{\ell m}(k).
\ee
We simplify this notation by defining $T_{\ell}(k;\Delta
r)=\int_{\Delta r} {\rm d}r j_{\ell}(kr)W(r)$ so that 
\be 
\bgamma^{2D}_{\ell m}(\Delta r)=\int {\rm d}k T_{\ell}(k;\Delta r)\bgamma_{\ell m}(k).
\ee
To find the power spectrum we take the covariance of both sides (using
the expressions from Castro et al., 2003 for the power spectrum) and find that 
\be
C^{2D}_{\ell}(\Delta_i r,\Delta_j r)=\int {\rm d}k_1 {\rm d}k_2
T_{\ell}(k_1;\Delta_i r)T_{\ell}(k_2;\Delta_j r) C_{\ell}(k_1,k_2),
\ee
where we now label a pair of bin-ranges in $r$ with $(i,j)$, and
$C_{\ell}(k_1,k_2)$ is the usual 3D cosmic shear power spectrum (or the pseudo power spectrum defined in Appendix B).

Using the Limber approximation, and with weight $W(r)=1$, the matrix $T$ becomes 
\be 
T_{\ell}(k;\Delta r=r_{\rm max}-r_{\rm min})\simeq
\left(\frac{\pi}{2\ell  k^2}\right)^{1/2} \,\,\,\, \forall
\,\,\, \frac{\ell}{r_{\rm max}}\leq k \leq \frac{\ell}{r_{\rm min}}
\ee
where the range of $\Delta r$ is explicit. For a single bin (all depth or 2D power spectrum) we have
$T^2_{\ell}(k)\simeq (\pi/2\ell k^2)$.

\newpage 
\section*{Appendix E: {\sc camb} priors} 
\begin{figure*}
\psfig{file=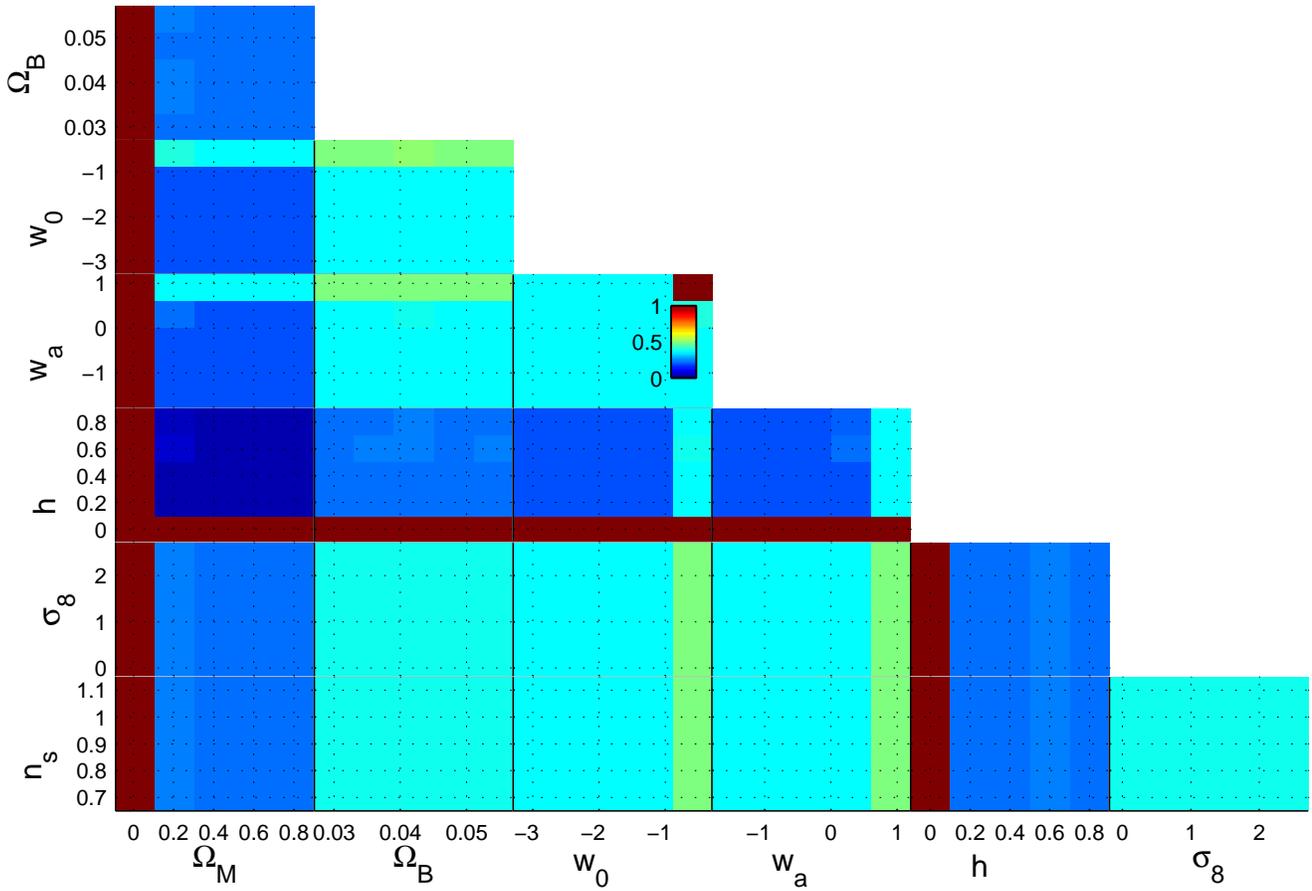, width=\columnwidth, angle=0}
\caption{The projected fraction of points from a $5^7$ grid that did not return a matter power spectrum from {\sc camb} (October 2012, with PPF module). The colour scale in the $(w_0,w_a)$ projected plane, that displays the colour associated with the fraction, is applicable to all other projections.}
\label{camb}
\end{figure*}

In this section we present a test of the software {\sc camb} used in this paper (October 2012, with the PPF module), which was used to justify 
some the prior ranges used. We sampled the waCDM parameter space on a grid containing $5^7$ points ($5$ in each parameter direction), 
in Figure \ref{camb} we show the projected fraction of these points 
that did not return a matter power spectrum. This functionality is reproduced by a very simple prior: $\Omega_{\rm M}<0.05 \vee h<0.1 \vee (w_0>-0.5 \wedge w_a>0.8)$. 


\newpage
\begin{thebibliography}{}

\bibitem[Albrecht et al.(2006)]{2006astro.ph..9591A} Albrecht, A., et al.\ 
2006, arXiv:astro-ph/0609591 

\bibitem[Amendola et al.(2013)]{2013LRR....16....6A} Amendola, L., et al.\ 2013, Living Reviews in Relativity, 16, 6 

\bibitem[]{}Anderson, T. W. 2003, An introduction to multivariate statistical analysis, 3rd edn. (Wiley-Interscience)

\bibitem[Ayaita et al.(2012)]{2012MNRAS.422.3056A} Ayaita, Y., Sch{\"a}fer, 
B.~M., \& Weber, M.\ 2012, \mnras, 422, 3056 

\bibitem[Ben{\'{\i}}tez(2000)]{2000ApJ...536..571B} Ben{\'{\i}}tez, N.\ 
2000, \apj, 536, 571 

\bibitem[Benjamin et al.(2013)]{2013MNRAS.431.1547B} Benjamin, J., Van 
Waerbeke, L., Heymans, C., et al.\ 2013, \mnras, 431, 1547

\bibitem[de Bernardis et al.(2009)]{2009PhRvD..80l3509D} de Bernardis, F., 
Kitching, T.~D., Heavens, A., \& Melchiorri, A.\ 2009, \prd, 80, 123509 

\bibitem[Brown et al.(2003)]{2003MNRAS.341..100B} Brown, M.~L., Taylor, 
A.~N., Bacon, D.~J., et al.\ 2003, \mnras, 341, 100 

\bibitem[Calabrese et al.(2013)]{2013arXiv1302.1841C} Calabrese, E., 
Hlozek, R.~A., Battaglia, N., et al.\ 2013, arXiv:1302.1841 

\bibitem[Camera et al.(2010)]{2010arXiv1002.4740C} Camera, S., Kitching, 
T.~D., Heavens, A.~F., Bertacca, D., \& Diaferio, A.\ 2010,
arXiv:1002.4740 

\bibitem[Castro et al.(2005)]{2005PhRvD..72b3516C} Castro, P.~G., Heavens, 
A.~F., \& Kitching, T.~D.\ 2005, \prd, 72, 023516 

\bibitem[Coles et al.(2004)]{2004MNRAS.350..989C} Coles, P., Dineen, P., 
Earl, J., \& Wright, D.\ 2004, \mnras, 350, 989

\bibitem[di Valentino et al.(2013)]{arXiv:1301.7343}
di Valentino E., et al.,  2013, arXiv:1301.7343

\bibitem[Driver \& Robotham(2010)]{2010MNRAS.407.2131D} Driver, S.~P., \& Robotham, A.~S.~G.\ 2010, \mnras, 407, 2131 

\bibitem[Erben et al.(2013)]{2013MNRAS.433.2545E} Erben, T., Hildebrandt, 
H., Miller, L., et al.\ 2013, \mnras, 433, 2545 

\bibitem[Fang et al.(2008)]{2008PhRvD..78h7303F} Fang, W., Hu, W., 
\& Lewis, A.\ 2008, \prd, 78, 087303 

\bibitem[Fang et al.(2008)]{2008PhRvD..78j3509F} Fang, W., Wang, S., Hu, 
W., et al.\ 2008, \prd, 78, 103509 

\bibitem[Feeney, Peiris, Verde (2013)]{arXiv:1302.0014}
Feeney S.M., Peiris H.V., Verde L., 2013, arXiv:1302.0014

\bibitem[Fu et al.(2014)]{2014arXiv1404.5469F} Fu, L., Kilbinger, M., \ 2014, arXiv:1404.5469 

\bibitem[Gelman A. \& Rubin D. (1992)]{} Gelman, A., Rubin, D, \ 1992, Statist. Sci., 7, 4, 457

\bibitem[Giocoli et al.(2010)]{2010MNRAS.408..300G} Giocoli, C., 
Bartelmann, M., Sheth, R.~K., \& Cacciato, M.\ 2010, \mnras, 408, 300

\bibitem[Harnois-D{\'e}raps et al.(2012)]{2012MNRAS.426.1262H} 
Harnois-D{\'e}raps, J., Vafaei, S., 
\& Van Waerbeke, L.\ 2012, \mnras, 426, 1262

\bibitem[]{}Hartlap J., Simon P., Schneider P., 2007, A\&A., 464, 399

\bibitem[Heavens(2003)]{2003MNRAS.343.1327H} Heavens, A.\ 2003, \mnras, 
343, 1327 
\bibitem[Heavens et al.(2006)]{2006MNRAS.373..105H} Heavens, A.~F., 
Kitching, T.~D., \& Taylor, A.~N.\ 2006, \mnras, 373, 105 

\bibitem[Heavens et al.(2007)]{2007MNRAS.380.1029H} Heavens, A.~F., 
Kitching, T.~D., \& Verde, L.\ 2007, \mnras, 380, 1029 

\bibitem[Heymans et al.(2005)]{2005MNRAS.361..160H} Heymans, C., Brown, 
M.~L., Barden, M., et al.\ 2005, \mnras, 361, 160 

\bibitem[Heymans et al.(2012)]{2012MNRAS.427..146H} Heymans, C., Van 
Waerbeke, L., Miller, L., et al.\ 2012, \mnras, 427, 146

\bibitem[Heymans et al. (2013)]{} Heymans, C. et al., \ 2013, accepted to MNRAS

\bibitem[Hicken et al.(2009)]{2009ApJ...700.1097H} Hicken, M., Wood-Vasey, 
W.~M., Blondin, S., et al.\ 2009, \apj, 700, 1097 

\bibitem[Hikage et al.(2011)]{2011MNRAS.412...65H} Hikage, C., Takada, M., 
Hamana, T., \& Spergel, D.\ 2011, \mnras, 412, 65

\bibitem[Hildebrandt et al.(2012)]{2012MNRAS.421.2355H} Hildebrandt, H., 
Erben, T., Kuijken, K., et al.\ 2012, \mnras, 421, 2355 

\bibitem[Hinshaw et al.(2013)]{2013ApJS..208...19H} Hinshaw, G., Larson, 
D., Komatsu, E., et al.\ 2013, \apjs, 208, 19 

\bibitem[Hirata 
\& Seljak(2004)]{2004PhRvD..70f3526H} Hirata, C.~M., \& Seljak, U.\ 2004, \prd, 70, 063526 

\bibitem[Hivon et al.(2002)]{2002ApJ...567....2H} Hivon, E., G{\'o}rski, 
K.~M., Netterfield, C.~B., et al.\ 2002, \apj, 567, 2 

\bibitem[Hou et al.(2012)]{2012arXiv1212.6267H} Hou, Z., Reichardt, C.~L., 
Story, K.~T., et al.\ 2012, arXiv:1212.6267 

\bibitem[Hu(1999)]{1999ApJ...522L..21H} Hu, W.\ 1999, \apjl, 522, L21 

\bibitem[Hu 
\& Sawicki(2007)]{2007PhRvD..76j4043H} Hu, W., \& Sawicki, I.\ 2007, \prd, 76, 104043

\bibitem[Jimenez et al.(2010)]{2010JCAP...05..035J} Jimenez, R., Kitching, 
T., Pe{\~n}a-Garay, C., \& Verde, L.\ 2010, \jcap, 5, 35 

\bibitem[Jing et al.(2006)]{2006ApJ...640L.119J} Jing, Y.~P., Zhang, P., 
Lin, W.~P., Gao, L., \& Springel, V.\ 2006, \apjl, 640, L119 

\bibitem[]{}Kaufman G.M., 1967, Some Bayesian Moment Formulae, Report No. 6710, Center for Operations Research and Econometrics, Catholic University of Louvain, Heverlee, Belgium

\bibitem[]{}Kendall, W. S.; Barndorff-Nielson, O.; and van Lieshout, M. C. Current Trends in Stochastic Geometry: Likelihood and Computation. Boca Raton, FL: CRC Press, 1998.

\bibitem[]{}Kenney, J. F. and Keeping, E. S. ``Kurtosis.'' -7.12 in Mathematics of Statistics, Pt. 1, 3rd ed. Princeton, NJ: Van Nostrand, pp. 102-103, 1962.

\bibitem[Kiessling et al.(2011)]{2011MNRAS.414.2235K} Kiessling, A., 
Heavens, A.~F., Taylor, A.~N., \& Joachimi, B.\ 2011, \mnras, 414, 2235 

\bibitem[Kilbinger et al.(2013)]{2013MNRAS.430.2200K} Kilbinger, M., Fu, 
L., Heymans, C., et al.\ 2013, \mnras, 430, 2200 

\bibitem[Kitching (2007)]{} Kitching, T.~D; PhD Thesis, University of Edinburgh \ 2007

\bibitem[Kitching 
\& Taylor(2011)]{2011MNRAS.416.1717K} Kitching, T.~D., \& Taylor, A.~N.\ 2011, \mnras, 416, 1717 

\bibitem[Kitching et al.(2012)]{2012MNRAS.423.3163K} Kitching, T.~D., et al.\ 2012, \mnras, 423, 3163

\bibitem[Kitching et al. (2011)]{2011MNRAS.413.2923K} Kitching, T.~D., 
Heavens, A.~F., \& Miller, L.\ 2011, \mnras, 413, 2923 

\bibitem[Kitching et al.(2007)]{2007MNRAS.376..771K} Kitching, T.~D., 
Heavens, A.~F., Taylor, A.~N., Brown, M.~L., Meisenheimer, K., Wolf, C., 
Gray, M.~E., \& Bacon, D.~J.\ 2007, \mnras, 376, 771 

\bibitem[Kitching et al.(2008)]{2008PhRvD..77j3008K} Kitching, T.~D., 
Heavens, A.~F., Verde, L., Serra, P., 
\& Melchiorri, A.\ 2008, \prd, 77, 103008 

\bibitem[Kitching et al.(2008)]{2008MNRAS.390..149K} Kitching, T.~D., 
Miller, L., Heymans, C.~E., van Waerbeke, L., 
\& Heavens, A.~F.\ 2008, \mnras, 390, 149 

\bibitem[Kirkman et al.(2003)]{2003ApJS..149....1K} Kirkman, D., Tytler, 
D., Suzuki, N., O'Meara, J.~M., \& Lubin, D.\ 2003, \apjs, 149, 1 

\bibitem[Komatsu et al.(2011)]{2011ApJS..192...18K} Komatsu, E., Smith, 
K.~M., Dunkley, J., et al.\ 2011, \apjs, 192, 18 

\bibitem[Kosowsky(1998)]{1998astro.ph..5173K} Kosowsky, A.\ 1998, 
arXiv:astro-ph/9805173 

\bibitem[Leistedt et 
al.(2012)]{2012A&A...540A..60L} Leistedt, B., Rassat, A., R{\'e}fr{\'e}gier, A., \& Starck, J.-L.\ 2012, \aap, 540, A60 

\bibitem[Loverde 
\& Afshordi(2008)]{2008PhRvD..78l3506L} Loverde, M., \& Afshordi, N.\ 2008, \prd, 78, 123506 

\bibitem[Ma et al.(2006)]{2006ApJ...636...21M} Ma, Z., Hu, W., 
\& Huterer, D.\ 2006, \apj, 636, 21

\bibitem[Merkel \& Sch{\"a}fer(2013)]{2013MNRAS.434.1808M} Merkel, P.~M., \& Sch{\"a}fer, B.~M.\ 2013, \mnras, 434, 1808 

\bibitem[Miller et al.(2013)]{2013MNRAS.429.2858M} Miller, L., Heymans, C., 
Kitching, T.~D., et al.\ 2013, \mnras, 429, 2858 

\bibitem[Mandelbaum et al.(2011)]{2011MNRAS.410..844M} Mandelbaum, R., 
Blake, C., Bridle, S., et al.\ 2011, \mnras, 410, 844

\bibitem[Miller et al.(2007)]{2007MNRAS.382..315M} Miller, L., Kitching, 
T.~D., Heymans, C., Heavens, A.~F., 
\& van Waerbeke, L.\ 2007, \mnras, 382, 315

\bibitem[Munshi et al.(2011)]{2011MNRAS.416.1629M} Munshi, D., Kitching, 
T., Heavens, A., \& Coles, P.\ 2011, \mnras, 416, 1629 

\bibitem[]{}Nesser, F. \& Massey, J., 1993, IEEE Transactions of Information Theory, 39, 4

\bibitem[Percival et al.(2010)]{2010MNRAS.401.2148P} Percival, W.~J., Reid, 
B.~A., Eisenstein, D.~J., et al.\ 2010, \mnras, 401, 2148 

\bibitem[Pen et al.(2002)]{2002ApJ...567...31P} Pen, U.-L., Van Waerbeke, 
L., \& Mellier, Y.\ 2002, \apj, 567, 31 

\bibitem[Pen et al.(2003)]{2003MNRAS.346..994P} Pen, U.-L., Lu, T., van 
Waerbeke, L., \& Mellier, Y.\ 2003, \mnras, 346, 994 

\bibitem[]{}Picinbono B., 1996, IEEE Transcations on Signal Processing, 44, 10

\bibitem[Planck Collaboration (2013)]{2013arXiv1303.5076P} Planck 
Collaboration, \ 2013, arXiv:1303.5076 
 
\bibitem[Riemer-Sorenson, Parkinson, Davis (2013)]{arXiv:1301.7102}
Riemer-Sorenson S., Parkinson D., Davis T., 2013, arXiv:1301.7102

\bibitem[Riess et al.(2009)]{2009ApJ...699..539R} Riess, A.~G., Macri, L., 
Casertano, S., et al.\ 2009, \apj, 699, 539 

\bibitem[Riess et al.(2011)]{2011ApJ...730..119R} Riess, A.~G., Macri, L., 
Casertano, S., et al.\ 2011, \apj, 730, 119

\bibitem[S{\'a}nchez et al.(2012)]{2012MNRAS.425..415S} S{\'a}nchez, A.~G., 
Sc{\'o}ccola, C.~G., Ross, A.~J., et al.\ 2012, \mnras, 425, 415 

\bibitem[Semboloni et al.(2011)]{2011MNRAS.417.2020S} Semboloni, E., 
Hoekstra, H., Schaye, J., van Daalen, M.~P., \& McCarthy, I.~G.\ 2011, \mnras, 417, 2020 

\bibitem[Semboloni et al.(2013)]{2013MNRAS.434..148S} Semboloni, E., 
Hoekstra, H., \& Schaye, J.\ 2013, \mnras, 434, 148 

\bibitem[Simpson et al.(2013)]{2013MNRAS.429.2249S} Simpson, F., Heymans, 
C., Parkinson, D., et al.\ 2013, \mnras, 429, 2249 

\bibitem[Smith et al.(2003)]{2003MNRAS.341.1311S} Smith, R.~E., et al.\ 
2003, \mnras, 341, 1311 

\bibitem[Takahashi et al.(2012)]{2012ApJ...761..152T} Takahashi, R., Sato, 
M., Nishimichi, T., Taruya, A., \& Oguri, M.\ 2012, \apj, 761, 152 

\bibitem[Taylor et al.(2013)]{2013MNRAS.432.1928T} Taylor, A., Joachimi, 
B., \& Kitching, T.\ 2013, \mnras, 432, 1928 

\bibitem[Valageas(2013)]{2013arXiv1306.6151V} Valageas, P.\ 2013, 
arXiv:1306.6151 

\bibitem[van Daalen et al.(2011)]{2011MNRAS.415.3649V} van Daalen, M.~P., 
Schaye, J., Booth, C.~M., \& Dalla Vecchia, C.\ 2011, \mnras, 415, 3649 

\bibitem[]{}van Waerbeke, L. et al., \ 2012

\bibitem[Verde(2007)]{2007arXiv0712.3028V} Verde, L.\ 2007, arXiv:0712.3028 

\bibitem[White(2004)]{2004APh....22..211W} White, M.\ 2004, Astroparticle 
Physics, 22, 211 

\bibitem[]{}Whittaker, E. T. and Watson, G. N. A Course in Modern Analysis, 4th ed. Cambridge, England: Cambridge University Press, 1990.

\bibitem[Wright(2006)]{2006PASP..118.1711W} Wright, E.~L.\ 2006, PASP, 
118, 1711 

\bibitem[Yang et al.(2013)]{2013PhRvD..87b3511Y} Yang, X., Kratochvil, 
J.~M., Huffenberger, K., Haiman, Z., \& May, M.\ 2013, \prd, 87, 023511

\bibitem[Zentner et al.(2008)]{2008PhRvD..77d3507Z} Zentner, A.~R., Rudd, 
D.~H., \& Hu, W.\ 2008, \prd, 77, 043507 

\bibitem[Zhan 
\& Knox(2004)]{2004ApJ...616L..75Z} Zhan, H., \& Knox, L.\ 2004, \apjl, 616, L75 

\end{thebibliography}
\end{document}